\documentclass[useAMS,usenatbib,asm]{mn2e}
\usepackage{graphicx,fleqn,times,color}
\newcommand{\tal}{\it et al. \rm}

\title[Rings and spirals in barred galaxies - I]{Rings and spirals in barred galaxies. I Building blocks. }
\author[Athanassoula, Romero-G\'omez and
  Masdemont]{E. Athanassoula$^1$, M. Romero-G\'omez$^1$,
  J.J. Masdemont$^2$ \\ 
$^1$Laboratoire d'Astrophysique de Marseille (LAM), UMR6110, 
CNRS/Universit\'e de Provence,\\
Technop\^ole de Marseille Etoile, 38 rue Fr\'ed\'eric Joliot Curie, 13388 Marseille C\'edex 20, France\\
$^2$I.E.E.C \& Dep. Mat. Aplicada I, Universitat Polit\`ecnica de
Catalunya, Diagonal 647, 08028 Barcelona, Spain\\
}
\date{Received }

\begin{document}

\maketitle

\begin{abstract}
In this paper we present building blocks which can explain the
formation and properties both of spirals and of  
inner and outer rings in barred galaxies. We first briefly summarise
the main results of the full theoretical description we have given
elsewhere, presenting them in a more physical way, aimed to an 
understanding without the requirement of extended knowledge of
dynamical systems or of orbital structure. We introduce in this manner
the notion of manifolds, which can be thought of as tubes guiding the
orbits. The dynamics of these manifolds can govern 
the properties of spirals and of inner and outer rings in barred
galaxies. We find that the bar strength affects how unstable the
$L_1$ and $L_2$ Lagrangian points are, the motion within the manifold
tubes and the time necessary for particles in a manifold to make a
complete turn around the galactic centre. 
We also show that the strength of the bar, or, to be more
precise, of the non-axisymmetric forcing at and somewhat beyond the
corotation region, determines the resulting morphology. Thus, less
strong bars give rise to $R_1$ rings or pseudorings, while
stronger bars drive $R_2$, $R_1R_2$ and spiral morphologies. We
examine the morphology as a function of the main parameters of the bar
and present descriptive two dimensional plots to that avail. We also
derive how the manifold morphologies and properties are modified if
the $L_1$ and $L_2$ Lagrangian points become stable. Finally, we discuss how
dissipation affects the manifold properties and compare the manifolds
in gas-like and in stellar cases. Comparison with observations,
as well as clear predictions to be tested by observations will be
given in an accompanying paper.    
\end{abstract}

\begin{keywords}
galaxies: kinematics and dynamics -- galaxies: spiral -- galaxies: structure -- stellar dynamics 
\end{keywords}

\section{Introduction}
\label{sec:intro}

Disc galaxies have a number of substructures, the most spectacular
ones being their rings and their spirals. Barred galaxies, in
particular, often have global spiral structure. They have two arms
that often start from the ends of the bar and wind outwards
covering a 
considerable region of the disc. In fact, global spiral structure is
found more often in barred than in non-barred galaxies
\citep[e.g.][]{ElmegreenElmegreen89}.

Rings also are often found in barred
galaxies. They come in three varieties: nuclear rings, which are small
and surround the nucleus, inner rings (r), which have the same size as the
bar and are slightly elongated along it, and outer rings (R), which are
considerably larger, with a major axis of the order of twice the bar
size. Depending on their orientation with respect to the bar, outer
rings are called $R_1$, when their major axis is perpendicular to the
bar major axis, $R_2$, when their major axis is along the bar major
axis, and $R_1R_2$, when they have a component parallel to the bar and
a component perpendicular to it. The latter are less frequent than the
previous varieties. Observational studies of their properties,
including their frequencies, shapes and orientations, have been made
by \citet{Buta95}. 

In order to understand the formation, evolution, and properties of 
any given structure it is essential to first understand its building
blocks, i.e. the orbits that constitute it. 
This was clearly demonstrated in the
case of bars, whose building blocks are closed periodic orbits
elongated along the bar, generally called x$_1$
(\citealt{ContopoulosPapayannopoulos80},  \citealt{AthBMP83}).  
The study of these building blocks provided answers to a number of
crucial questions, like why bars are bisymmetric, why they 
rotate as rigid bodies, why they can not extend beyond corotation, why
peanuts and boxy bulges form and what their structure and extent should be,
etc. (\citealt{Contopoulos81}, \citealt{Binney81}, \citealt{Pfenniger84},
\citealt{SkokosPA02a}, \citealt{PatsisSA02}, \citealt{Ath05}, etc.).
Since bars are present in most disc galaxies, such studies went a long
way towards explaining not only bar properties, but also bar
formation and evolution, as well as the evolution of disc galaxies in
general. 

Spirals and rings, both inner and outer, also are present in a large
fraction of disc galaxies. Identifying their building blocks will help
explaining their formation and evolution, as well as their
properties. It could also give information on the properties of the
underlying disc galaxy and on the pattern speed and relative strength of
the bar. For this purpose, we need to find the building blocks of spiral
arms and of inner and outer rings in barred galaxies and to study their
properties. Of course such a study is only a step towards
understanding a given structure, since it neglects collective effects
which can play an important role. Yet it can provide a full physical
understanding. It is this first step that is the 
aim of our work. We started it in two previous papers
(\citealt{RomeroGMAG06}, hereafter  
Paper I, and \citealt{RomeroGAMG07}, hereafter Paper II), while a more
analytical approach to the problem can be found in \citet{RomeroGMGA08}. 

In Papers I and II
we proposed a theory which can explain the formation of both rings and
spirals in barred galaxies using a common framework. It is based on the
chaotic orbital motion driven by the unstable equilibrium points of
the rotating bar potential. We thus suggested that spirals, rings and
pseudorings are 
related to the invariant manifolds associated to the periodic orbits
around these equilibrium points and, particularly, to the existence of
heteroclinic or homoclinic orbits. Thus, $R_1$ rings are associated
to the presence of heteroclinic orbits, while $R_1R_2$ are associated
to the presence of homoclinic orbits. Spiral arms and $R_2$ rings,
however, are present when there exist neither heteroclinic nor
homoclinic orbits. To establish this link, we calculated both
manifolds and their associated trajectories in a large number of
cases, covering the relevant parameter space of three   
simple barred galaxy models. This allowed us to discuss the
formation of different morphological structures according to the
properties of the galaxy models. 
Work along similar lines has also been carried out by other
teams. \cite{Danby65} argued that bar orbits departing from the
vicinity of the unstable Lagrangian points play an important role in
the formation of the spiral arms. \citet{KaufmannContopoulos96} 
linked chaotic 
orbits to the presence of spiral arms. This was further developed by 
\citet{Patsis06} with the help of response calculations in the
potential of NGC 4314, as calculated by \citet{QuillenPG94}, and by
\cite{VoglisSK06} who used a potential from an $N$-body simulation.
Manifolds were specifically referred to, although in a quite
different way from that used in our work, by the late Prof. Voglis and his
collaborators \citep{VoglisTE06}.  

The above cited works were crucial in establishing the importance of
chaos in the formation of spirals and rings in barred galaxies and,
more specifically, the role of the manifolds.
Yet a considerable amount of work still needs to be done, particularly
in the practical aspects, since a number of crucial questions are
still unanswered. Which properties of the manifolds influence most
those of rings and spirals? Which properties of the galactic potential
determine whether the morphology will be that of a ring or that of a
spiral? In other words, is it possible from the barred galaxy
potential to predict the galaxy morphology? If yes, this would be a
clear prediction and therefore a test of our theory. How does the
morphology of the manifolds compare with those of observed spirals and
rings? Are such 
manifolds to be always expected, or are there cases where the unstable
Lagrangian points can somehow become stable? What happens to the manifolds
in such cases? We attempt to answer these
questions in this paper (Paper III of this series).

In section \ref{sec:theory} we give a simplified and physical
description of the main theoretical tools necessary for this
study. We describe here the Lagrangian points, the invariant manifolds
and their associated orbits. Section
\ref{sec:originmorpho} links manifold morphologies to 
bar properties. Section~\ref{sec:manfprop} introduces
some useful properties of the invariant manifolds. 
Section~\ref{sec:L12stability} addresses the option of stable
$L_1$ and $L_2$ Lagrangian points and discusses the type of
morphologies this would entail. Section~\ref{sec:schwarz} discusses
gas and compares its response morphology to that of manifolds. 
We briefly summarise in section~\ref{sec:summary}. Comparison with
observations will be presented in Paper IV of this series, where we will
also give a global discussion on the applicability of our
theory and a brief comparison with other spiral structure theories. 

\section{Theoretical basis}
\label{sec:theory}

In this section we summarise the main theoretical tools necessary for 
our study. Our description aims towards a physical understanding and
is thus wilfully somewhat simplified. A more accurate and thorough
description is given in Papers I and II and in the references therein,
while some analytical 
intricacies of manifolds in a simplified, rotating, non-axisymmetric
potential have been presented by \citet{RomeroGMGA08}. 

\subsection{General}
\label{subsec:gentheory}

As in the two previous papers, we use here simple, rigid models, since these
contain all the basic, necessary physics. They are composed of an
axisymmetric part and of a bar rotating with a constant angular velocity,
which we will refer to as the pattern speed. 
We concentrate on the motion on the $z$ = 0
 plane (equatorial plane of the galaxy), since the motion in the
vertical direction can be essentially described by an uncoupled 
harmonic oscillator, is assumed to be of relatively small amplitude
and does not affect the motion in the $z$ = 0 plane 
(Paper I). We also limit ourselves to bars rotating in the direct
sense, i.e. here counter-clockwise, in agreement with all
observational and simulation results. We will work in a frame of
reference rotating with the bar, i.e. in which the bar is at rest,
since the dynamics are much easier described there
\citep{BinneyTremaine08}. We use the 
convention that, in this frame, the bar is along the $x$ axis. 

We consider three different barred galaxy models,
described in Appendix~\ref{app:models}. They consist of an
axisymmetric component and a rigid, non-evolving bar, rotating with
a constant angular velocity $\Omega_p$. Model A
has a Ferrers' bar \citep{Ferrers77}, 
characterised by its semi-major axis $a$, its axial ratio
$a/b$ and its quadrupole moment $Q_m$. Our fiducial case all through
this paper is a typical example of a model with a Ferrers' bar
and has $a=5$, $a/b=2.5$, $r_l=6$, $Q_m=4.5 \times 10^4$,
$\rho_c=2.4 \times 10^4$ 
and $n=1$, all in the units given in Appendix~\ref{app:models}. Here
$r_L$ is the Lagrangian radius, $\rho_c$ is the central concentration
of the system and $n$ is the index of the Ferrers' bar
(Appendix~\ref{app:models}). Our conclusions, however, 
are based on a large number of such models, covering a wide range of
parameters. 

Models D and BW have ad hoc formulae for the bar potential and 
were initially given by 
\citeauthor{Dehnen00} (\citeyear{Dehnen00}) and
by \citeauthor{BarbanisWoltjer67} (\citeyear{BarbanisWoltjer67}),
respectively. Their bar strength is 
quantified with the free parameters $\epsilon$ and $\hat{\epsilon}$,
respectively. They can 
have a much stronger non-axisymmetric forcing beyond corotation than
models A, and are meant to represent and include forcings not only from
bars, but also from spirals, from oval discs and from triaxial
haloes. We have kept all these forcings bar-like, i.e. we
have not included any radial variation of the azimuthal dependence.
This was done on purpose, in
order not to bias the response towards spirals and in order to
avoid the extra degree of freedom resulting from the azimuthal
winding of the force.     

\subsection{Equilibrium points and Lyapunov orbits}
\label{subsec:Lagratheory}

\begin{figure}
\begin{center}
\includegraphics[scale=0.4,angle=-90.0]{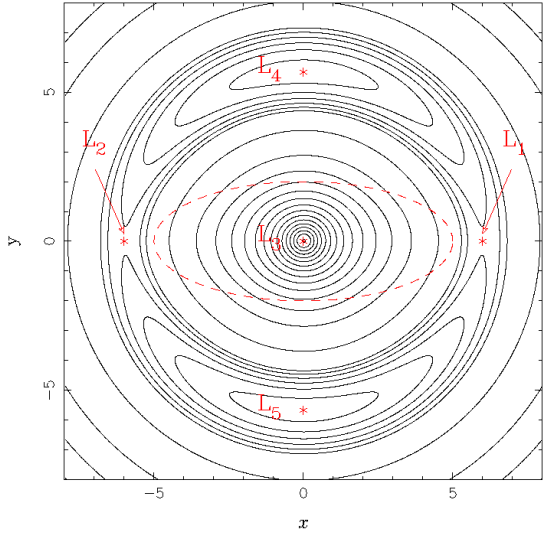}
\end{center}
\caption{Contours of constant effective potential. The values of the
  iso-effective potentials have been chosen so as to show best the
  relevant dynamical features. The outline of
  the bar is given by a dashed line and the five Lagrangian
points are marked by asterisks. } 
\label{fig:effpot1}
\end{figure}

For barred galaxy models such as those used here, the energy of a
particle in the rotating frame is a constant of the motion and is
often referred to as the Jacobi constant, $E_J$, or simply as the
energy \citep{BinneyTremaine08}. A useful quantity
to define here is the effective potential, 

$$\Phi_{\hbox{\scriptsize eff}}=\Phi-\frac{1}{2}\Omega_p^2\,(x^2+y^2),
$$

\noindent
which can be thought of as the potential in the rotating frame of
reference. $\Omega_p$ is the bar pattern speed, i.e. the angular
velocity with which the bar rotates, assumed constant, and $\Phi$ is
the potential. The curve $\Phi_{\hbox{\scriptsize eff}}=E_J$ is called
the zero velocity curve (ZVC). All regions in which $\Phi_{\hbox{\scriptsize
eff}}>E_J$ are forbidden to particles with Jacobi constant equal to
$E_J$ or less, and are therefore called forbidden regions.

The motion has five equilibrium points, i.e five points where 

$$\frac{\partial \Phi_{\hbox{\scriptsize eff}}}{\partial x}=
\frac{\partial \Phi_{\hbox{\scriptsize eff}}}{\partial y}= 0.$$ 

\noindent
They 
are often called the Lagrangian points $L_i$, $i$ = 1, .. 5 and
their location is shown in Fig.~\ref{fig:effpot1}. $L_3$ is at the
origin of the coordinates, $L_4$ and $L_5$ lie on the direction of the
bar minor axis, symmetrically with respect to the centre, and $L_1$
and $L_2$ lie on the direction of the bar major axis, also
symmetrically with respect to the centre. For realistic bar
potentials, the distance of the $L_4$ and $L_5$ from the centre is
somewhat larger than that of the $L_1$ and $L_2$ \citep{Ath92a}. 
We will refer to the distance of the $L_1$ and $L_2$ from the centre
as the Lagrangian radius $r_L$. The $L_4$ and $L_5$ are maxima
of the effective potential, $L_3$ is a minimum and $L_1$ and $L_2$ are saddle
points, i.e. there
$$
\frac{\partial^2 \Phi_{\hbox{\scriptsize eff}}}{\partial x^2}<0,~~
\frac{\partial^2 \Phi_{\hbox{\scriptsize eff}}}{\partial y^2}>0.
$$ 

\begin{figure}
\begin{center}
\includegraphics[scale=0.35,angle=-90.0]{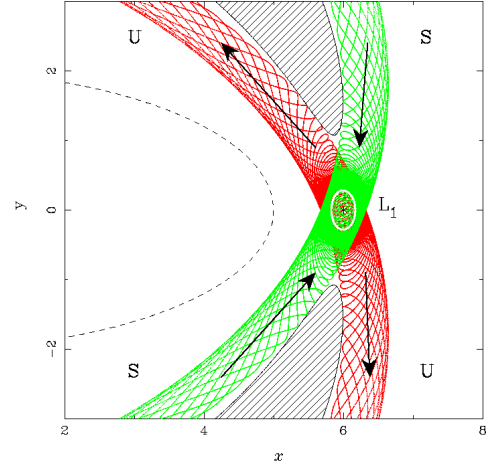}
\end{center}
\caption{Dynamics of the region around $L_1$ for a typical value of
  the energy. The location of $L_1$ is given by an asterisk and we
  show in a white solid line the corresponding Lyapunov orbit around it. 
  The two branches of the unstable invariant manifold (marked $U$ and
  plotted with red lines), and the two branches of the stable
  invariant manifold (marked $S$ and plotted with green lines) are for
  the same value of the energy. The  
  arrows give the direction of the motion along the manifold and the
  hatched areas are the forbidden regions surrounded by 
  the zero velocity curves (solid black lines). The loci of the
  manifolds in the vicinity of $L_2$ is
  identical, but mirrored with respect to the bar minor axis. The
  outline of the bar is given by a black dashed line. All through this paper
  and wherever a distinction is necessary, the unstable manifolds will
  be plotted in red and the stable ones in green. 
} 
\label{fig:branches}
\end{figure}

$L_3$ is stable and is
surrounded by the x$_1$ family of periodic orbits
\citep{ContopoulosPapayannopoulos80}, which are the backbone of the
bar \citep{AthBMP83}. $L_4$ and $L_5$ are also generally stable and each
one of them is 
surrounded by a family of periodic orbits, called the banana 
orbits because of their shape. 
These three Lagrangian points and the corresponding families
of periodic orbits have been studied extensively \citep[see][and
  references therein]{Contopoulos02}. $L_1$ and $L_2$ are generally unstable
and are surrounded by 
a family of periodic orbits, often called the Lyapunov orbits
\citep{Lyapunov49}. These orbits have a roughly elliptical shape, their
size increases with energy and they are
unstable, becoming stable only at energies much higher 
than that of $L_{1}$ and $L_{2}$ \citep{SkokosPA02a}. An example is
  shown in Fig.~\ref{fig:branches}. The dynamics of 
the $L_{1}$ and $L_{2}$ points and of the Lyapunov family 
of periodic orbits had been little studied before we started on
Paper I, presumably 
because, being unstable, they were deemed less interesting. Yet, as we
showed in Papers I and II and will further stress here, the
manifolds they generate can account for the spirals, as well as for the
inner and outer rings observed in barred spirals. 

Since the Lyapunov orbits are unstable, they can not trap around them
any regular orbits, so that any orbit with initial conditions in their
vicinity in phase space will not stay near them. The time it
takes for the orbit to leave the vicinity (in the phase space sense) of
the corresponding Lyapunov periodic orbit depends on its Jacobi
constant. 
For lower values of the energy, i.e. values near that of $L_1$, the
Lyapunov orbits have a smaller extent and are more unstable. Thus,
any orbit starting from their immediate vicinity in phase space will
escape their neighbourhood quite fast, in a time corresponding to a
couple of bar rotations.
For larger values of the energy, the Lyapunov orbits become larger
and more asymmetric and they are less unstable, so that the orbits
starting from their immediate vicinity in phase space will take
longer to escape.

\subsection{Invariant manifolds and associated orbits}
\label{subsec:manifolds}

\begin{figure*}
\centering
\includegraphics[scale=0.5,angle=-90.]{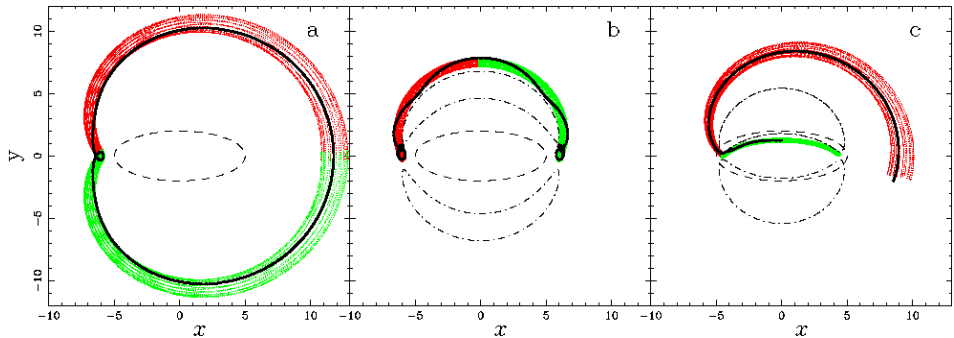}
\caption{Examples of homoclinic {\bf (a)}, heteroclinic {\bf (b)} and
  escaping 
{\bf (c)} orbits (black thick lines) in the configuration space. In
red, we plot the unstable invariant manifolds associated to these
orbits, while in green we plot the corresponding stable invariant
manifolds. In dashed lines, we give the outline of the bar and, in 
{\bf (b)} and {\bf (c)}, we plot in dot-dashed lines the zero velocity
curves of the same energy as the orbits and manifolds.   
}
\label{fig:homoheterosp}
\end{figure*}

Since any orbit in the immediate vicinity (in phase space) of the
unstable Lyapunov orbits can not stay trapped around them,
it will have to escape the neighbourhood of the corresponding
Lagrangian point. Not all departure directions are, however, 
possible. Let us consider a Lyapunov orbit of a given energy and
another orbit of the same energy and initially very close in 
phase space to the Lyapunov one. The direction in which this orbit
escapes is set by what 
is called the invariant manifolds. We can simply think
of the manifolds as tubes which guide the orbits escaping from the
vicinity of the Lagrangian points, so that these manifolds/tubes are
filled and surrounded by orbits. We refer the reader to Paper I and
to the references therein for a precise definition and for a description
of how we calculate the manifolds in practise. 

Fig.~\ref{fig:branches} explains the dynamics around the $L_1$ 
Lagrangian point for a typical value of the energy. We plot here (in
white) the Lyapunov orbit of that energy and the four branches of the
invariant manifold that emanate from it, two inner and two outer. For
two of them, one inner and one outer, the motion is away from the
region of the Lyapunov orbit and they are referred to in the theory of
dynamical systems as the unstable branches of  
the manifold. For the other two -- again one inner and one outer -- the
motion is towards the Lyapunov orbit and they are referred to in the
theory of dynamical systems as the stable branches of
the manifold. We will use these terms here also, but we want to stress
that this does {\it not} mean that the orbits that are guided by these
branches are also stable or unstable, respectively. In fact these
orbits are all chaotic, but they are in a loose way `confined' by the
manifolds, so that they stay together in what could be described as a
bundle, at least for a couple of rotations around the bar. In that
sense, the manifolds can be thought of as driving the 
dynamics in the vicinity of the $L_1$ and $L_2$,  

Fig.~\ref{fig:branches} shows only the vicinity of the Lagrangian
point, but, as can be seen in Fig.~\ref{fig:homoheterosp}, the manifolds
extend far beyond this neighbourhood. They can thus be
responsible for more global structures in the galaxy. It should, however,
be stressed that not all particles can be affected by the manifold
dynamics, but only those in a relatively narrow energy range, whose lower
limit, as described in Paper I, corresponds to the energy of
the $L_1$ and $L_2$ points. Since two of the manifold branches 
have motions inwards and two outwards, manifolds can
play a crucial role in the transport of material between different parts 
of the galaxy. Loosely, they can be thought 
of as gates between the regions within and the regions outside
corotation (as in astrodynamics, see Koon \tal 2000, Gomez \tal 2004).

The morphology of the manifolds with respect to the Lagrangian points allows
us to classify their outer branches in three types, homoclinic,
heteroclinic and escaping. This is illustrated in 
Fig.~\ref{fig:homoheterosp}, where we show, for each case, the manifolds
and a corresponding orbit. All depart from an unstable Lyapunov
periodic orbit around one of the unstable equilibrium points, in this
case $L_2$. We define as homoclinic the manifolds and orbits that
return to it (Fig. \ref{fig:homoheterosp}a). Similarly, heteroclinic
manifolds and orbits are those that approach the 
corresponding Lyapunov periodic orbit around the Lagrangian point at
the opposite end of the bar, $L_1$ (Fig. \ref{fig:homoheterosp}b). 
Finally, there are manifolds and orbits that do not
return to either $L_1$ or $L_2$, but spiral outwards from the region
of the unstable Lyapunov periodic orbits to reach the outer regions of the
galaxy. Following the notation of Paper II, we refer to them as escaping.
This means that they can reach regions far from the vicinity
of the bar, but {\it not} that they can escape to infinity. They
correspond to the unstable branches of the manifold and the motion
along them is outwards and in the clockwise (retrograde) sense. 
Similarly to the escaping manifolds and orbits, there are incoming
manifolds and orbits, which, coming from the outer parts of the galaxy,
reach the vicinity of $L_1$ or $L_2$. They correspond to the stable
branches of the manifold and their loci can be obtained from the unstable
ones after a reflection with respect to the bar major axis. The motion
along them is inwards and anticlockwise (direct). These, however, as
we will show in Paper IV,
do not have any physical significance
in our problem, so we will not discuss them much.

There is a further crucial difference between the homoclinic and
heteroclinic cases on the one hand, and the escaping ones on the
other. In the former the stable and unstable branches overlap, at
least partly, both in configuration space (positions) and in phase
space (position, velocity space). The contrary is true for the
latter. Thus, in the former cases orbits can be trapped (guided) by
both the stable and the unstable branches of the manifold
simultaneously. This is, for example, the case for the two orbits 
plotted by solid black lines in the left and middle panels of
Fig.~\ref{fig:homoheterosp}. This, however, is not the case for the
orbit in the right panel, which belongs to the unstable manifold
exclusively. In this sense, the red and green colours in the left and
middle panels are arbitrary, since any part of the manifold/orbit can
be assigned to both the stable and the unstable branches. 

We will argue here that the three types of orbits we presented here --
i.e. the homoclinic, the 
heteroclinic, and the escaping orbits -- are the backbone of ringed
structures and of spiral arms observed in disc galaxies, in the same
way as x$_1$ orbits are the backbone of the bar. 

\begin{figure*}
\centering
\hspace{0.5cm}
\includegraphics[scale=0.7,angle=-90.0]{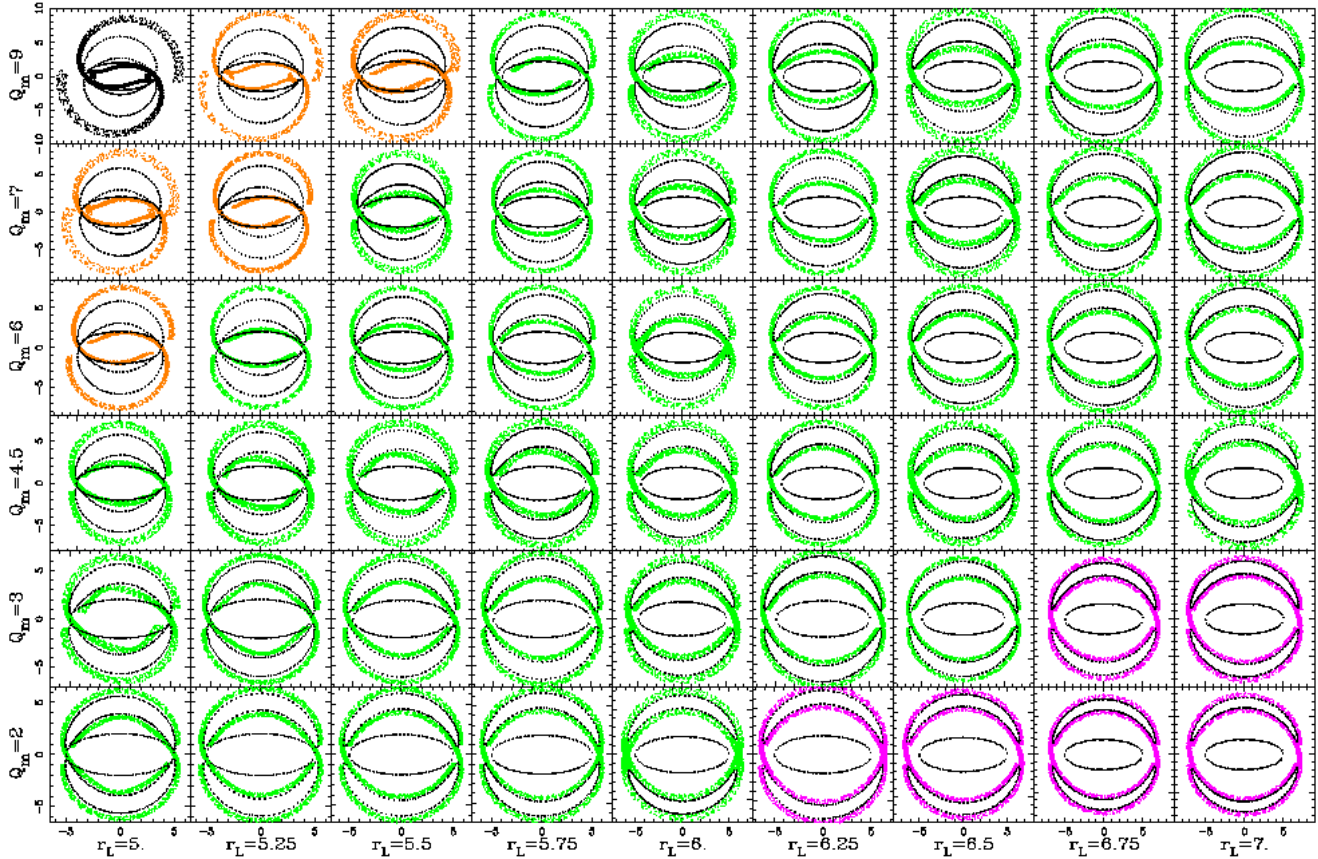} \hspace{0.5cm}
\caption{Effect of the bar strength and of the Lagrangian radius $r_L$
  on the manifold loci for model A. The strength of the bar is
  parametrised by the quadrupole moment $Q_m$.
  Each panel corresponds to a pair of ($Q_m$, $r_L$) values, given on
  the left (in units of 10$^4$) and on the bottom of the figure,
  respectively. The remaining parameters are as in the fiducial
  model. The 
  manifolds are plotted in a colour which is determined by their
  morphology: green for $R_1$, orange for $R_1'$ and black for
  spirals (see text for 
  lilac). The black dotted lines give the outline of the bar and the
  zero velocity curves of the same energy as the manifolds.}
\label{fig:2Dferrers}
\end{figure*}

Two more points need to be made here. One is that the manifolds and
orbits we present here are only the building blocks, but do not in
any way assure us that the corresponding structure will be present
in the galaxy. Indeed a given manifold may exist, but may not trap
any orbits, so that the corresponding structure will not be
visible in the galaxy. This is similar to the periodic orbits, which
are the building blocks of bars and which need to trap regular orbits
around them, for the bar to form. In other words, the existence of a
manifold is a necessary, but not a sufficient condition for the
corresponding galactic structure to form.

The second point we wish to stress is that the orbits guided by the
manifolds are not the only ones in the relevant galactic
regions. On the contrary, there are a number of other possible
orbits. For example, there are families of periodic orbits beyond
corotation (e.g. the x$_1'$ in the notation of
\cite{ContopoulosPapayannopoulos80} and the A', B', -2/1 and -1/1 in
the notation of \cite{AthBMP83}) that can have large stable
parts so that the corresponding periodic orbits can trap regular orbits
around them. It is the ensemble of these orbits, those driven by the
manifolds and those trapped around the stable periodic orbits, that
will structure the region beyond corotation. Here we concentrate on
the manifold-driven orbits in order to propose them as possible
building blocks for spirals and rings.
   
\section{manifold morphologies and bar properties}
\label{sec:originmorpho}

\begin{figure*}
\centering
\hspace{0.5cm}
\includegraphics[scale=0.7,angle=-90.0]{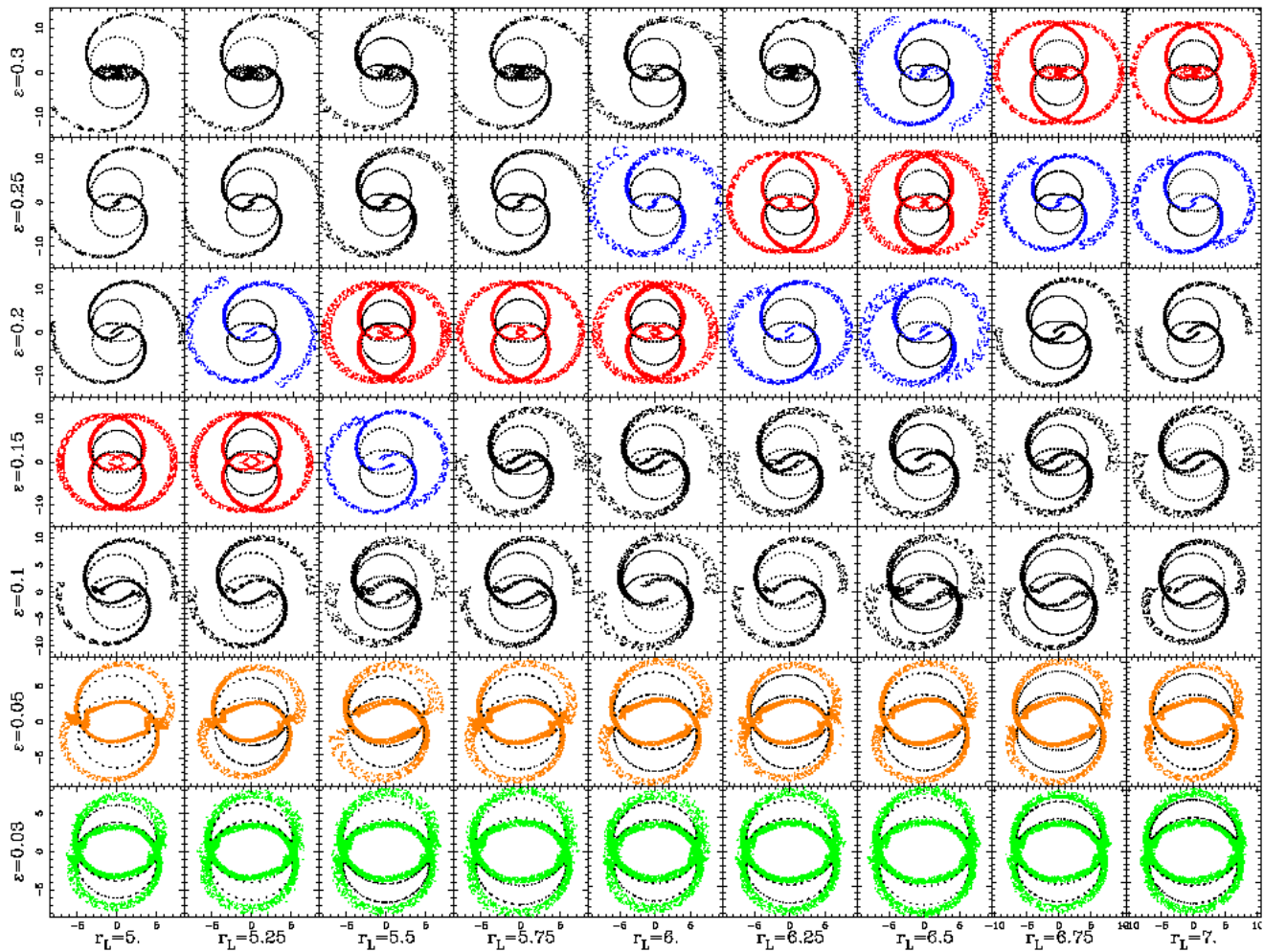} \hspace{0.5cm}
\caption{As for Fig.~\ref{fig:2Dferrers}, but for D type models.
  The parameter $\epsilon$ is a measure of the bar strength. $R_2$
  rings are plotted in blue and $R_1R_2$ in red. 
}
\label{fig:2Ddehnen}
\end{figure*}

\begin{figure*}
\centering
\hspace{0.5cm}
\includegraphics[scale=0.7,angle=-90.0]{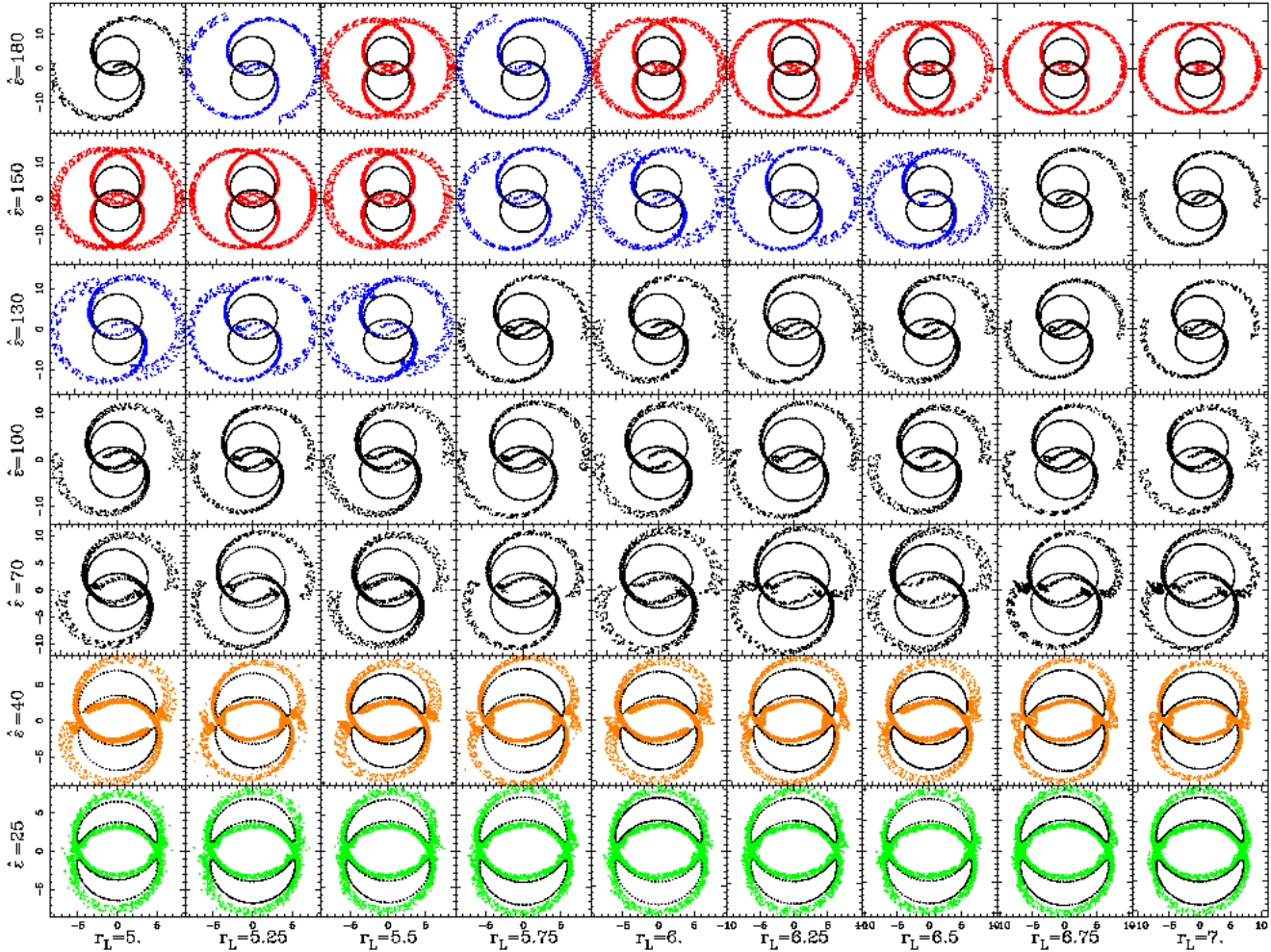} \hspace{0.5cm}
\caption{As for Fig.~\ref{fig:2Ddehnen}, but for BW
  type models. The parameter $\hat\epsilon$ is a measure of the bar
  strength. 
}
\label{fig:2Dbw}
\end{figure*}

How does the shape of the outer manifold loci depend on the parameters 
determining the bar potential? To answer this question we calculated
manifolds in a very large number of models from all three families
of potentials and include some of the manifold loci in
Figs.~\ref{fig:2Dferrers} to \ref{fig:2Dbw}.
Fig.~\ref{fig:2Dferrers} corresponds to
models of type A, Fig.~\ref{fig:2Ddehnen} to models of type D and
Fig.~\ref{fig:2Dbw} to models of type BW. 

Models of type A have 4 main free parameters : their
central concentration $\rho_c$, their bar axial ratio $a/b$, their  
quadrupole moment $Q_m$ and their pattern speed parametrised by the
value of the Lagrangian radius $r_L$. As we showed in Paper II, 
two of these, the central density $\rho_c$ and the axial ratio $a/b$,
influence the potential mainly within corotation, so that we do not
need to consider their effect in depth here. The other two,
$Q_m$ and $r_L$, influence strongly the manifold loci also
beyond corotation, so we will mainly limit our discussion to
them. Contrary to Paper II, where we only presented the manifold
shapes for barred galaxy models on the axes of this
parameter space, we will consider here a grid of relevant values for
these parameters. In this way we can get a much better insight of
the dependence of the manifold morphology on the bar properties.
The other two families of models, D and BW, have only two free
parameters, the first one ($\epsilon$ or $\hat{\epsilon}$,
respectively) linked to the bar strength, and the second one
($r_L$) being the Lagrangian radius. We will thus for each of the models
present a composite two-dimensional plot including many models.
Each row
corresponds to a given value of the bar strength, and each column to a
given value of the pattern speed. In fact, both of these parameters
influence the strength of the non-axisymmetric forcing at and somewhat
beyond the corotation radius.

We classified the morphologies of the outer manifold branches into
spirals (black), $R_1$ rings (light green), $R_1$ pseudorings
(orange), $R_2$ rings and pseudorings (blue) and $R_1R_2$ 
rings and pseudorings (red). For model A, we plot in lilac cases with
$rR_1$ morphology, but which have a
ratio of inner to outer major axes which does not seem compatible with
observations. The ones corresponding to the  the smallest bar strengths
have two rings with diameters which do not differ much. In cases where
the bar amplitude is so low, the motion in this region may be almost
independent of the bar, i.e. governed more or less by the axisymmetric
mass distribution. In
such cases, once self-gravity is taken into account, the two rings
could merge, giving rise to a thicker, single, very low amplitude feature. 

How is this classification done? For the models, it is possible to use 
the presence of homoclinic, heteroclinic and escaping manifolds and orbits
in order to classify the morphologies. This, however, has not been done
by observers classifying real galaxies, who use eye
classification. Since we wish to make 
extensive comparisons with observations in Paper IV,
we will use the same classification means as observers. In general, it
is easy to differentiate by eye between the different types of
morphology; borderlines cases, however, are 
a matter of personal judgement. There are such borderline cases
between $R_1$ and $R_1'$, between $R_1'$ and spirals, etc. This should
be taken into account when assessing any of the results obtained here
and when discussing statistical results in Paper IV.

Figures~\ref{fig:2Dferrers} to \ref{fig:2Dbw} reveal a very clear
trend. Namely the different 
morphologies are not randomly distributed in the (strength, Lagrangian
radius) plane, but, on the contrary, different morphologies are grouped
together in different parts of that plane, which, in a very rough way,
can be seen as 
diagonal stripes. $R_1$ rings occupy the bottom right, followed by
spirals, then by $R_2$ and $R_1R_2$ and finally by more $R_2$ and
spirals.

This clearly indicates that the morphology depends on the strength of
the non-axisymmetric forcing 
in the part of the galaxy occupied by the manifolds. Indeed, the bar
strength increases from the bottom to the top row. The pattern speed,
however, also influences the bar strength in that region. As the
pattern speed decreases (i.e. as the Lagrangian radius $r_L$
increases) the distance between the end of the bar and the $L_1$ (or
$L_2$) increases, so that the manifolds will occupy a region farther
away from the bar, i.e. a region which has a weaker non-axisymmetric
forcing. Thus, the influence of the bar is stronger as we move towards
stronger bars (bottom to top in each figure) and as the pattern speed
increases (right to left in each figure).
 
In the language of dynamical systems, Figs.~\ref{fig:2Dferrers} to
\ref{fig:2Dbw} show that the strength 
of the perturbation in the region of the manifolds determines whether
these are homoclinic, heteroclinic, or of escaping type 
(Fig.~\ref{fig:homoheterosp}). Alternatively, in 
more astronomical terms, these plots argue that the strength of the
bar in the region beyond but close to corotation should 
determine whether we will have spirals or rings, and more
specifically what type of rings.

\begin{figure}
\centering
\hspace{0.5cm}
\includegraphics[scale=0.4,angle=-90.0]{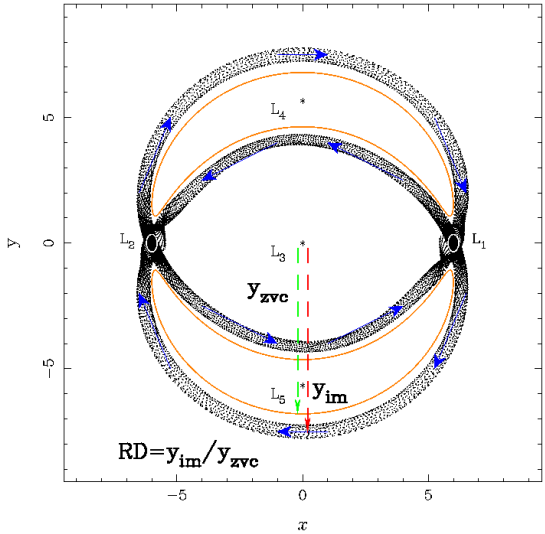} \hspace{0.5cm}
\caption{Example of a manifold with $rR_1$ morphology (black). The
sense of circulation of material 
along it is shown by blue arrows. The ZVC of the same energy is shown
in orange. The distance from the centre of the galaxy  to the outer
part of the ZVC ($y_{ZVC}$) and the distance
from the centre of the galaxy  to the outer branch of the manifold
($y_{im}$), both measured along the $y$ axis, are marked by dashed
arrows in green and red, respectively. Their ratio,
$RD=y_{im}/y_{ZVC}$, is used in Fig.~\ref{fig:Ratdist-qt} to delimit
the various morphologies.} 
\label{fig:RDcirculation}
\end{figure}

To distinguish further between the different types of responses we
introduce two relevant quantities. The first one quantifies the
relative bar strength in the region of $L_1$.
For this we use a standard measure of the bar strength at a given
radius, namely 

\begin{equation}
Q_t (r) = (\partial \Phi (r, \theta) /\partial \theta)_{max}/(r\partial
\Phi_0/\partial r),
\label{eq:Qt}
\end{equation}

\noindent
where $\Phi$ is the potential, $\Phi_0$ is its axisymmetric part and the
maximum in the numerator is calculated over all values of the
azimuthal angle 
$\theta$. We calculate $Q_t (r)$ at the radius of $L_1$, i.e. at
$r = r_{L1}$ and denote it by $Q_{t,L_1}$.
The choice of the second relevant quantity is motivated by the results in
Sect.~\ref{sec:manfprop}. More specifically, we will show in that section
that, in the cases where $L_1$ and $L_2$ are weakly unstable, the loci
of the 
manifolds are located very near the zero velocity curve of the same
energy, while, in the strongly unstable case, they depart considerably
from it. We quantify this by measuring on the $y$ axis the ratio of
two distances, namely the distance between the centre of the galaxy and
the manifold ($y_{im}$ in Fig.~\ref{fig:RDcirculation}) and the distance 
between the centre and the outer branch of the corresponding ZVC
($y_{ZVC}$ in Fig.~\ref{fig:RDcirculation}). We refer to this ratio
($y_{im}$/$y_{ZVC}$) as $RD$. As will be shown in
Sect.~\ref{sec:manfprop}, both quantities $Q_{t,L_1}$ and $RD$ measure
the bar strength, albeit in a different way. 

\begin{figure}
\centering
\hspace{0.5cm}
\includegraphics[scale=0.35,angle=-90.0]{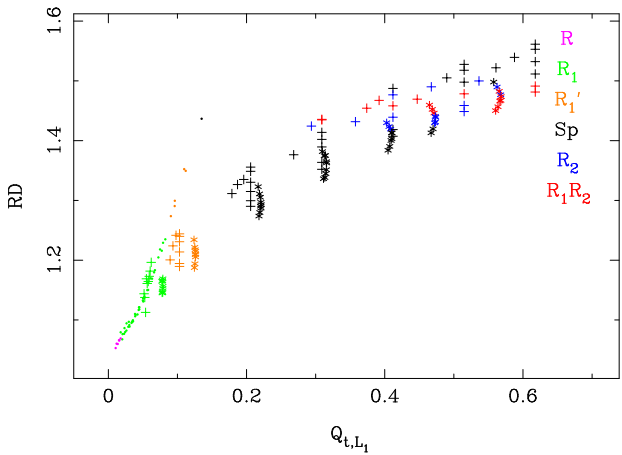} \hspace{0.5cm}
\caption{Location of the models shown in Figs~\ref{fig:2Dferrers}
  to \ref{fig:2Dbw} in the ($RD$, $Q_{t,L_1}$) plane. $RD$ is the
  ratio of $y_{im}$ (the distance from the centre of the galaxy to the
  outer branch of the manifold) to $y_{ZVC}$ 
  (the distance from the centre to the branch of the corresponding
  ZVC farthest from the centre). $Q_{t,L_1}$ is a measure of the bar
  strength 
  at the radius of the $L_1$, given by eq.~(\ref{eq:Qt}). The various
  morphologies -- $R_1$, $R_1'$, spirals, $R_2$ and $R_1R_2$ -- are
  noted by symbols of different colour,
  as given in the figure. The various symbols show whether
  the model is A (dots), D (crosses), or BW (asterisks). 
}
\label{fig:Ratdist-qt}
\end{figure}

With the help of these two quantities we can include the main
information in a single plot, given in Fig.~\ref{fig:Ratdist-qt}. This
shows the location of all models from Figs~\ref{fig:2Dferrers} to
\ref{fig:2Dbw} on the ($Q_{t,L_1}$, $RD$) plane and 
reveals that the different types of morphologies are segregated on
different parts of this plane, independent of the type of the
model. For the lowest $Q_{t,L_1}$ values (roughly $Q_{t,L_1} < 
0.1$) we have only $R_1$ outer rings, while somewhat higher values 
(roughly in the region $0.1 < Q_{t,L_1} < 0.2$) give $R_1'$ 
morphologies. Yet higher values give spirals and $R_2$ and $R_1R_2$ rings. 
Fig.~\ref{fig:Ratdist-qt} shows that there is no division between
these three types by $Q_{t,L_1}$ value alone, but that there is one,
albeit a bit rough, by the value of $RD$. 

Following the location of the models on the ($Q_{t,L_1}$, $RD$)
plane, after the $R_1$ and the $R_1'$ we find spirals, followed by
$R_2$ rings and pseudorings, then $R_1R_2$, then again $R_2$ and again
spirals. This second group of spirals has much more open arms than the
first one. 

We redid this plot, using, instead of the value of $Q_{t,L_1}$,
the average value of $Q_t$ in an annulus of given width starting at
$L_1$ and extending beyond it and found qualitatively the same results for a
wide range of widths. It is important to underline that this plot
includes three different types of bar potentials, which are widely
different. Thus,
Fig.~\ref{fig:Ratdist-qt} argues that the morphological segregation
exists for many, if not all, reasonable bar potentials. It also shows,
together with Figs.~\ref{fig:2Dferrers} to \ref{fig:2Dbw}, that a
spiral forcing is not mandatory in order to obtain a spiral
response, but that a spiral can be obtained with a purely bar or
bar-like forcing,
provided of course this is sufficiently strong in the regions beyond
corotation. This has been already shown to be the case for the density
wave theory \citep[e.g.][]{FeldmanLin73,Ath80}. It is nevertheless true
that it is easier to obtain a spiral response with a spiral than a
with a bar forcing. It is also clear that if one adds a spiral forcing
to the bar one, it will be even easier to obtain spirals, as
underlined already by \cite{LindbladLA96}, in their modelling of NGC
1365. Thus, for bar plus spiral non-axisymmetric forcings spirals will
form for lower relative strength than for the bar-only case. Therefore, 
the exact delimitation between spirals and rings in
e.g. Fig.~\ref{fig:Ratdist-qt} will depend on whether the
forcing is bar only, or bar plus spiral.

\section{Further properties of the invariant manifolds}
\label{sec:manfprop}

We saw in Sect.~\ref{subsec:Lagratheory} that the Lagrangian points $L_1$
and $L_2$ are generally 
unstable, but we still need to quantify how unstable they are. For this we 
use a positive quantity $\lambda$, introduced and defined in appendix B.
The larger numerical values of $\lambda$ correspond to more unstable 
$L_1$ and $L_2$ Lagrangian points. 

\begin{figure}
\centering
\includegraphics[scale=0.33,angle=-90.0]{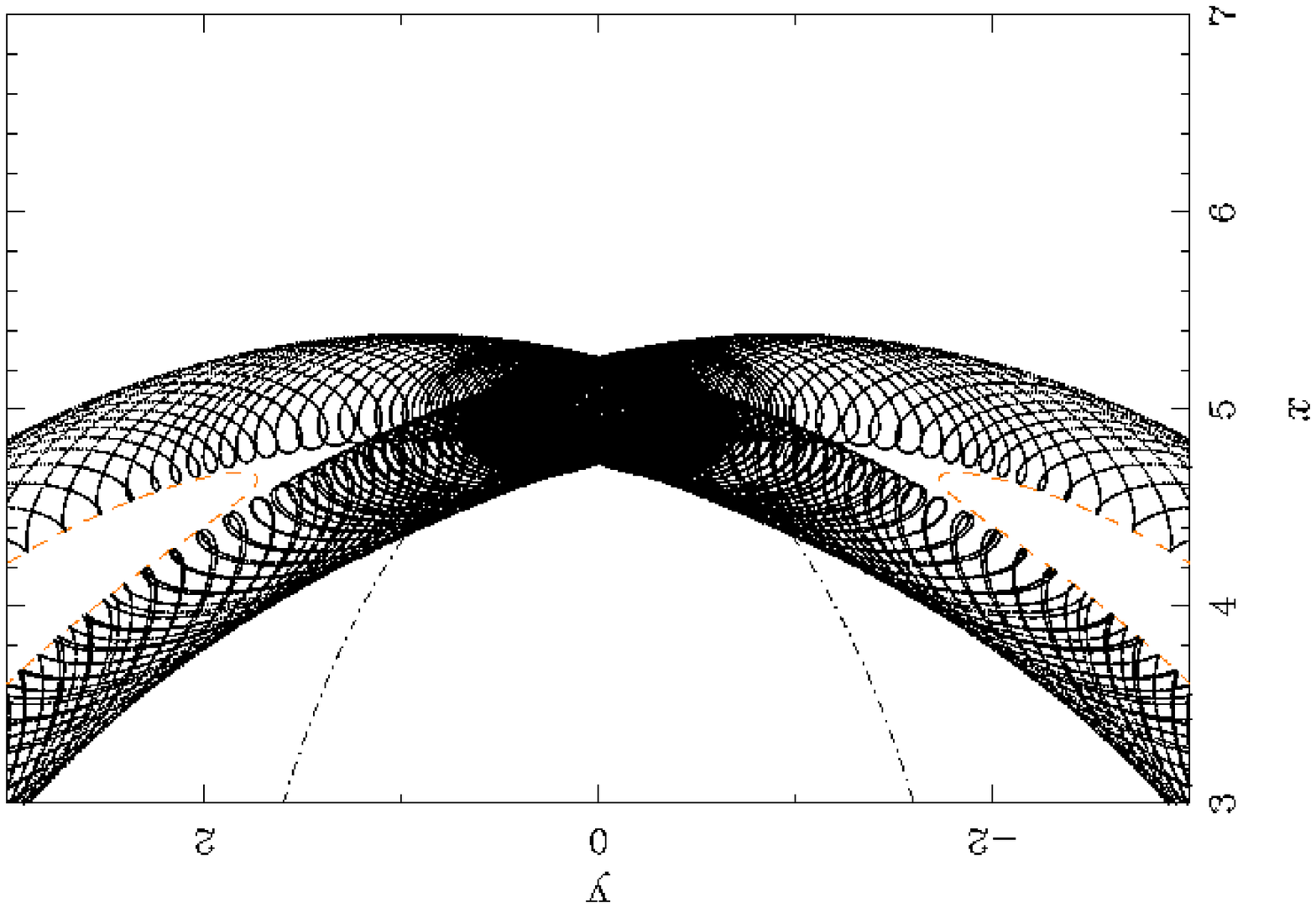}
\includegraphics[scale=0.33,angle=-90.0]{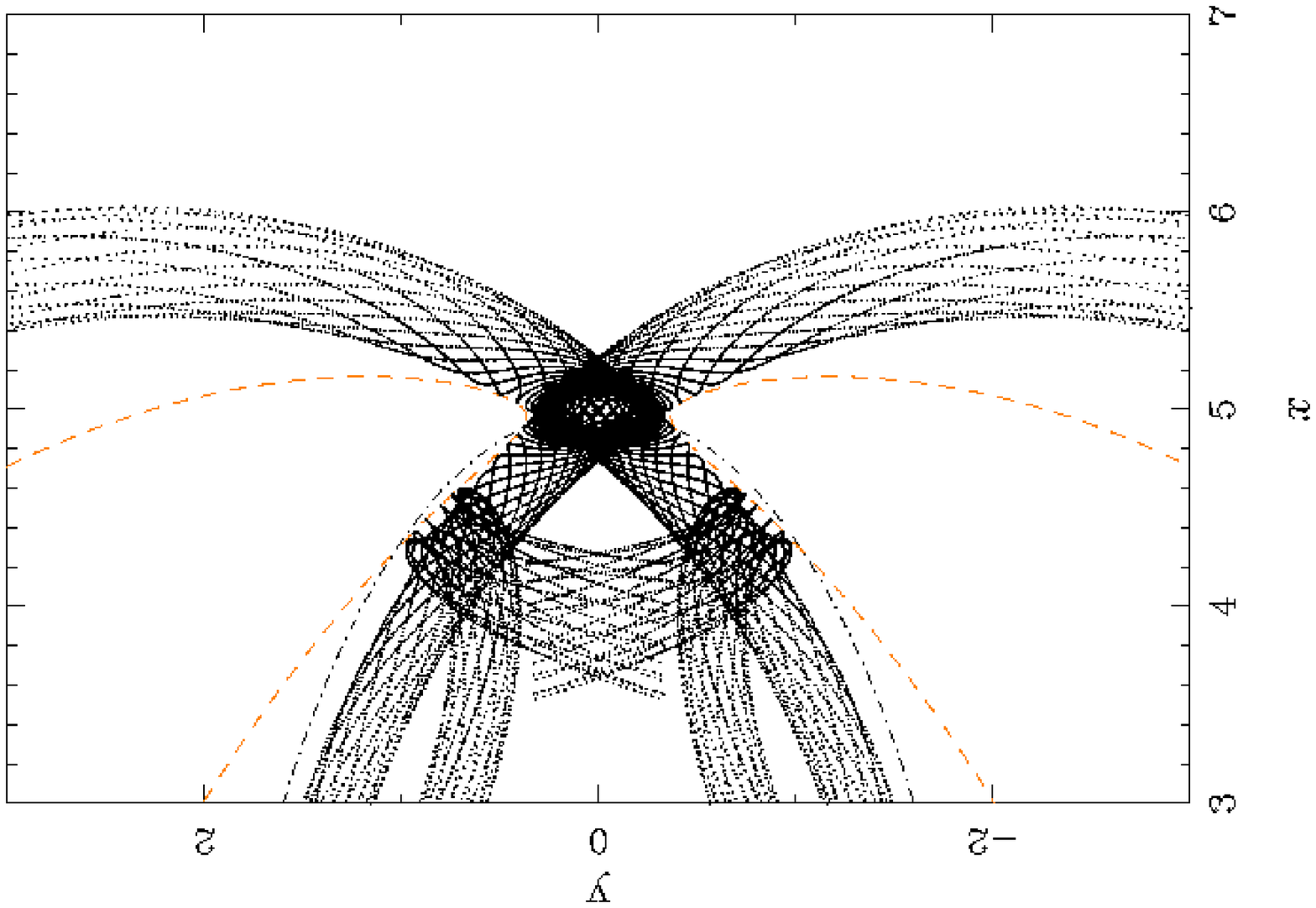}
\caption{Invariant manifolds in the immediate neighbourhood
of $L_1$ for a strongly unstable case (right panel) and for a weakly
unstable one (left panel). More than one revolution is shown for the
inner manifold branches. The dot-dashed black lines show the bar outline
and the the dashed orange ones the zero velocity curves.
}
\label{fig:weakstrman}
\end{figure}

Fig.~\ref{fig:weakstrman} shows the invariant manifolds in the
immediate neighbourhood of $L_1$ for a strongly unstable case
and for a weakly unstable one. We chose as examples a model
with a strong bar ($Q_m$ = 9, right panel of
Fig.~\ref{fig:weakstrman}) and with a weak bar ($Q_m$ = 0.5, left panel of 
Fig.~\ref{fig:weakstrman}), both with $r_L$ = 5. These models have 
$\lambda$ = 62.85 and $\lambda$ = 11.88, respectively. There are
several clear differences, revealing how the strength of the bar
influences the instability of the $L_1$ and $L_2$
and the manifold properties. As can be seen in
Fig.~\ref{fig:weakstrman}, in the weakly unstable case (left panel) a
considerable part of the motion is perpendicular to the main extent of
the manifold, so that the manifolds are more densely packed and $T_f$, the
time 
necessary for them to perform half a revolution around the galaxy
centre, is very long. On the contrary,
for a strongly unstable case (right panel) most of the motion is along
the main extent of the manifolds, so that the manifolds are less
densely packed and $T_f$ is much shorter. 

\begin{figure}
\centering
\includegraphics[scale=0.25,angle=-90.0]{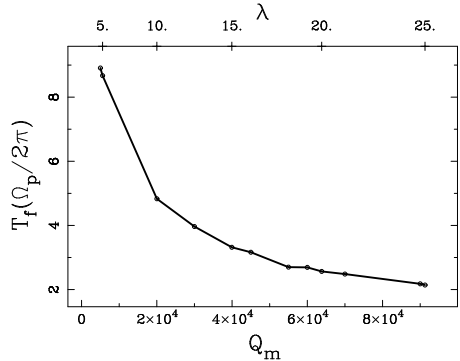}
\caption{$T_f$, i.e. the time necessary for the outer branch of a
  manifold starting from the 
  immediate neighbourhood (in phase space) of a Lyapunov periodic
  orbit to perform half a revolution around the galactic centre, as a
  function of $Q_m$ (the quadrupole moment of the bar, lower abscissa)
  and of the $\lambda$ parameter  measuring the instability of the
  Lagrangian point (upper abscissa). $T_f$ is measured in units of the
  bar rotation period and $\Omega_p$ is the bar pattern speed. This
  figure shows clearly that strong 
  bars (i.e.bars with large $Q_m$ values) have more unstable $L_1$
  ($L_2$) Lagrangian points (i.e. larger $\lambda$ values) and develop
  their manifolds in shorter times. 
}
\label{fig:Tf}
\end{figure}

This is illustrated in Fig.~\ref{fig:Tf}, which shows the time $T_f$
necessary for the outer branch of a manifold starting from the
immediate neighbourhood (in phase space, see Paper I) of a Lyapunov
periodic orbit to perform half a 
revolution around the galactic centre, as a function of the
quadrupole moment, $Q_m$, of the bar\footnote{Models with different
  $Q_m$ values have different values of the potential energy at the
  Lagrangian point. To make the comparison in Fig.~\ref{fig:Tf} fair,
  we compare manifolds with the same energy relative to that
  of the $L_1$ of their model. Thus the energy of all the manifolds is
  $E_J(L_1) + \delta\epsilon$, where $E_J(L_1)$ is the potential energy at
  $L_1$ of the corresponding model and $\delta\epsilon$ is a very
  small shift, the same for all models. In this example, we chose
  $\delta\epsilon$ = 20.5, but the resulting behaviour is independent
  of the $\delta\epsilon$ value chosen. Note also that these times 
refer to the outer branches of the manifolds.}. The remaining parameters are
those of the fiducial A model (Sect.~\ref{subsec:gentheory}). 
Calculating the corresponding $\lambda$ values for each of the barred
potentials, we see that, as expected, stronger bars make more unstable
Lagrangian points. We place the corresponding values of
$\lambda$ on the upper limit of the plot, to show how $\lambda$
depends on $Q_m$ and also how $T_f$ depends on $\lambda$. Fig.~\ref{fig:Tf}
shows clearly that $T_f$ drops very fast with increasing bar
strength, in an exponential-like way. 

This is not the only difference. As can be seen in
Fig.~\ref{fig:weakstrman}, the loci of the manifolds in the
weakly unstable case are located very near the zero velocity curve
of the same energy, contrary to the manifolds of the strongly unstable
case, which are at a considerable distance from it. Moreover,
Figs.~\ref{fig:2Dferrers} to \ref{fig:2Dbw} show that
this is true not only in the vicinity of $L_1$, or $L_2$, but also
over all their extent. Thus, after a rotation of 90$^\circ$ the
manifolds of the weakly unstable cases are much
nearer to the centre than those of the strongly unstable cases. This
conditions the shape of the manifold loci and explains the result 
found in Sect.~\ref{sec:originmorpho}. 

\begin{figure}
\centering
\includegraphics[scale=0.25,angle=-90.0]{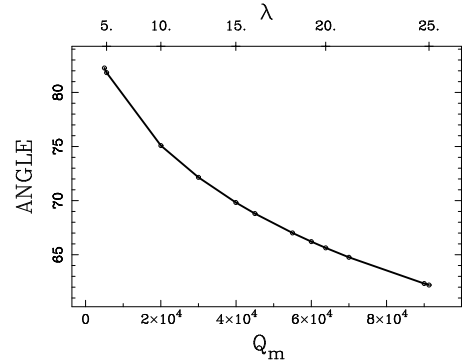}
\caption{Angle between the direction of the manifold near the
  Lagrangian point and the direction of the bar major axis, plotted as a
  function of $Q_m$ (the quadrupole moment of the bar, lower abscissa)
  and of the $\lambda$ parameter measuring the instability of the
  Lagrangian point (upper abscissa). Strongly unstable configurations
  have large values of $\lambda$ and smaller values of the angle.
} 
\label{fig:manifangle}
\end{figure}

Another way of seeing this is by measuring the angle between the
direction of the outer manifold branches near $L_1$ and the direction
of the bar major 
axis. For weakly unstable cases -- as in the left panel of
Fig.~\ref{fig:weakstrman} -- this angle is large, tending to
$90^{\circ}$ as the $L_1$ and $L_2$ become less and less unstable. On
the contrary, this angle is smaller in the case of strongly unstable
Lagrangian points. This is illustrated in Fig.~\ref{fig:manifangle}, 
which plots this angle as a function of $Q_m$ (lower abscissa) and
$\lambda$ (upper abscissa), for the same models as in Fig.~\ref{fig:Tf}.

There are yet further differences. 
Weak bars drive particles on heteroclinic orbits (see
Sect.~\ref{subsec:manifolds}) and produce the shape of $rR_1$ rings
and pseudorings
(see Sect.~\ref{sec:originmorpho}). The circulation of the material
within the manifolds is shown in Fig.~\ref{fig:RDcirculation}. Four
different paths are possible. Material can circulate along the inner
branches of the manifold (i.e. keeping within the inner ring), or
along the outer manifold branches (i.e. keeping within the outer
ring), or have a mixed trajectory. The latter is particularly
interesting. Mass circulates within the
manifolds, moving from $L_1$ along the inner ring (inner branches of the
manifold) to the vicinity of the $L_2$ and from there
outwards on the outer ring (i.e. on an outer branch of the manifold) until
it reaches a maximum distance from the centre. At this point the radial
component of the velocity changes sign and mass elements will move
inwards, still keeping on the outer branch of the manifold, back towards the
Lagrangian point $L_1$. This closes one complete circulation path. This
path and the direction of motion along it are shown in 
Fig.~\ref{fig:RDcirculation}. There are two such paths, one above and
one below the $x$ axis, symmetric with respect to the bar major
axis. The four circulation paths can be repeated or alternated. For
all four, there is no net inwards or outwards 
motion. Material, however, circulates within the galaxy, so that there
is a radial, as well as azimuthal mixing. In particular, for the paths
involving both inner and outer branches, material circulates from the
region within corotation to the region outside it and vice versa.
Initially, matter is trapped in these 
manifolds at the time the bar is formed, while further material is
added as the bar strength increases.  

For very strongly unstable $L_1$ and $L_2$, the situation is different. 
Material now follows escaping orbits, which trace the spirals. 
The stable and unstable manifolds intersect in the configuration
space [the ($x$, $y$) space], but not in phase space [i.e. not in the
(position, velocity) space]. Indeed, on the $y$ axis, where the stable
and unstable branches intersect in the configuration space, the radial
velocity component of the particles 
driven by the unstable branch is outwards, while that of
particles on the stable branch is inwards (note that at this point
the radial component of the velocity was equal to zero for the
heteroclinic case). Thus, particles
initially trapped in the inner branches of the
manifold will move towards one of the unstable Lagrangian points, say
the $L_2$, and from there outwards guided by an unstable outer branch of the
manifold. Contrary to the case of homoclinic, or heteroclinic
manifolds and orbits, they will not come back to $L_1$ or $L_2$,  
Thus the motion along first the inner stable manifold and then
the outer unstable manifold will take mass from the outer region of
the bar, which is a high density region in the galaxy, and feed it
into the spiral. This will bring a redistribution of matter within the
galaxy and a net outwards motion. Of course, in principle,
material could also come inwards 
starting from the outer parts of the disc on the stable outer branches
of the manifolds and reach the outer parts of the bar. This would
imply leading spirals bringing material inwards towards the outer
parts of the bar. In Paper IV 
we will describe the dynamical reasons that disfavour this possibility
and the effect of these different types of circulation on galactic discs.

A further important difference concerns the inner branches of the
manifolds. If we consider integration times longer than $T_f$, i.e. if
we trace the manifolds over more than half a revolution around the
galactic centre, then in the weakly unstable case the manifolds
retrace the same loci, forming an inner ring. In other words, there is
an overlap between the stable and the unstable inner manifolds, in the sense
described in Sect.~\ref{subsec:manifolds}. 
This is not the case for the strongly unstable cases, where the inner
manifolds cover a new path after the first revolution around the
galactic centre and retrace the same loci only after a couple or a few
revolutions.  

\section{Stabilisation of $L_1$ and $L_2$ and related manifold formation}
\label{sec:L12stability}

\begin{figure}
\begin{center}
\includegraphics[scale=0.45,angle=-90.0]{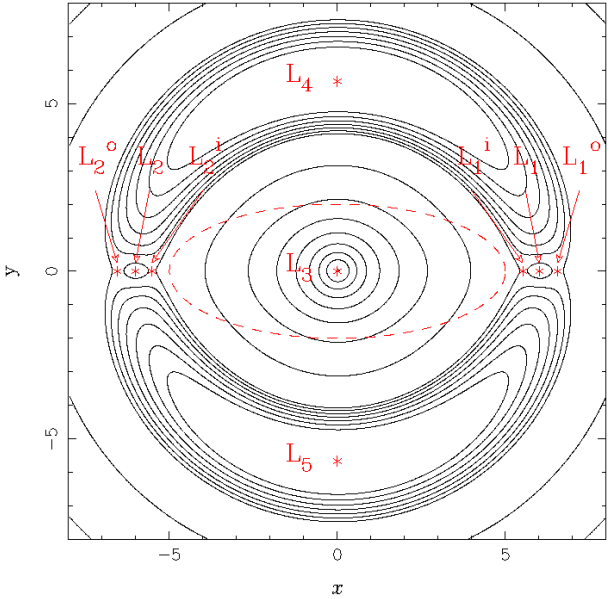}
\end{center}
\caption{Contours of constant effective potential. The model is the
  same as in Fig.~\ref{fig:effpot1}, except that we have added two small
  Kuz'min/Toomre discs centred on the positions of the $L_1$ and $L_2$ (see
  text). The total mass of each of these discs is $M_a$=0.025$M_{b}$
  and their scale length is $r_a$ = 0.6. All Lagrangian
  points are marked by asterisks. Note that there are now nine 
  Lagrangian points. The $L_{1}$ and $L_{2}$ are now
  minima of the effective potential and each one has on either side
  of it two more Lagrangian points. These are called $L^i_{1}$ (or
  $L^i_{2}$) and $L^o_1$ (or $L^o_2$), for the inner and outer one,
  respectively, and are saddle points.  } 
\label{fig:effpot2}
\end{figure}


In order for the theory as described so far to be
applicable to barred spirals, their Lagrangian points $L_1$ and $L_2$
must be unstable. We will hereafter refer to this case as the standard
case. If this were not the case, i.e. if the $L_1$ and $L_2$ became somehow
stable, the Lyapunov orbits would also be stable even very near the
$L_1$ and $L_2$ and thus the invariant manifolds described in
Sect.~\ref{subsec:manifolds} would not form. We thus want to test in
this section whether  
the $L_1$ and $L_2$ are always unstable, or whether they
could become stable, and, if so, under what conditions. 

One way of achieving this stability could be by adding a
concentration of matter around the $L_1$ and $L_2$. We tested this
by adding two identical, small Kuz'min/Toomre discs
\citep{Kuzmin56,Toomre63}, one centred on each of the 
$L_{1}$ and $L_{2}$. As can be seen in Fig.~\ref{fig:effpot2}, this
changes the topology of the iso-effective-potential curves. The
$L_{1}$ and $L_{2}$, instead of being saddle points as in
the standard case (Fig.~\ref{fig:effpot1}), become minima. On either
side of them 
two more equilibrium points form\footnote{In the language of dynamical
systems, the two new equilibrium points bifurcate from the $L_1$ (or
$L_2$).}, one at larger and the other at smaller radii, 
which we call $L^i_{j}$ and $L^o_j$ ($j = 1, 2$), respectively for the
inner and outer one. 
These are saddle points, so we can expect them to influence the
dynamics in a way similar to that of the unstable $L_{1}$ ($L_{2}$) in
the standard case.

Where could such an extra mass around the Lagrangian points come from?
As we saw in Sect.~\ref{subsec:Lagratheory}, any orbit in the
immediate vicinity of a Lyapunov 
orbit will not leave it immediately, but will first circle around it a
couple or a few times before following the direction of the
manifold. This would result in a mass concentration around $L_1$
($L_2$) which would contribute to the necessary mass
concentration. Furthermore, mass in this region could be contributed by
specific morphological characteristics of the bar, such as star
formation near the ends of the bar, or ansae
  \citep{Sandage61, Ath84, MartinezVKB07}. This issue will be further
  discussed in Paper IV.  

\begin{figure}
\centering
\includegraphics[scale=0.3,angle=-90.0]{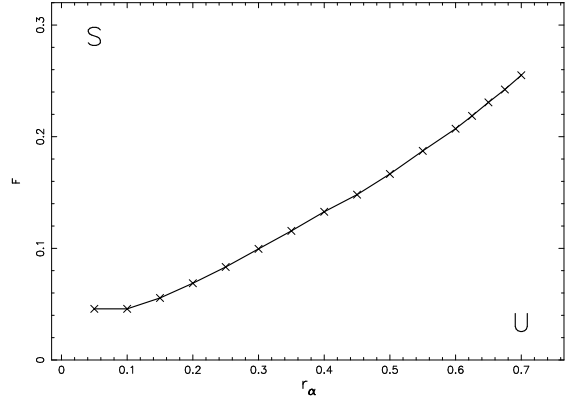}
\caption{Stability diagram for the models discussed in
  Sect.~\ref{sec:L12stability}. The solid line separates the region
  within which the $L_1$ and $L_2$ are stable from the 
  region in which they are unstable; the stable being in the upper left
  part. $F$ is a measure of the mass of the Kuz'min/Toomre disc stabilising
  the $L_1$, or $L_2$ (see text) and $r_a$ is its scale length. }
\label{fig:eigenval}
\end{figure}

We studied the stability of the $L_1$ and $L_2$ using the method
described in Appendix~\ref{app:stability}  and in Sect. 3 of
Paper I. Two quantities influence this 
stability, namely the mass and the scale length of the two Kuz'min/Toomre
discs, assumed to be identical. Since normalised quantities are more
intuitive, instead of the mass we use the quantity
$F=M_{KT}(r<r_a)/M_{b}(r<r_a)$, where $r_a$ is the scale length of the
Kuz'min/Toomre disc, $M_{KT}(r<r_a)$ is its mass within a distance equal to
$r_a$ from its centre and $M_{b}(r<r_a)$ is the mass of the bar within
the same distance from the galaxy centre. 
The results of this stability analysis are shown in
Fig.~\ref{fig:eigenval}. The solid line separates the stable from the
unstable region; the stable one being in the upper left part. As
expected, stability is achieved when the Kuz'min/Toomre discs
superposed on the $L_1$ and $L_2$ are sufficiently massive and 
sufficiently concentrated, in which case 
the $L_1$ and $L_2$ will become stable. 

\begin{figure}
\begin{center}
\includegraphics[scale=0.45,angle=-90.0]{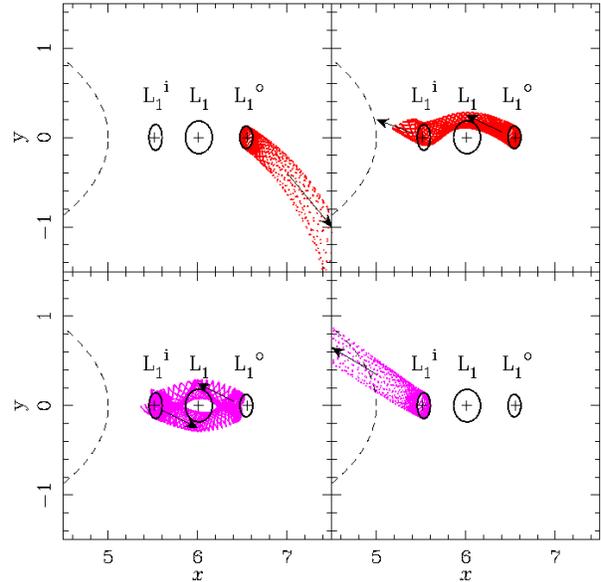}
\end{center}
\caption{Unstable manifolds for a case with a stable $L_1$ Lagrangian
  point. The model is the same as for Fig.~\ref{fig:effpot2}. On
  either side of the $L_1$ we note the inner and outer 
  unstable Lagrangian 
  points, $L_1^i$ and $L_1^o$, respectively, each surrounded by a
  periodic orbit. The unstable manifolds for the same energy are also
  plotted (red). The upper (lower) panels show the unstable manifolds of the
  outer (inner) Lagrangian point, $L^o_1$ ($L^i_1$). The arrows show
  the direction of the motion within the manifolds and the dashed line
  gives the outline of the bar.}
\label{fig:3Lunstable}
\end{figure}

\begin{figure}
\begin{center}
\includegraphics[scale=0.45,angle=-90.0]{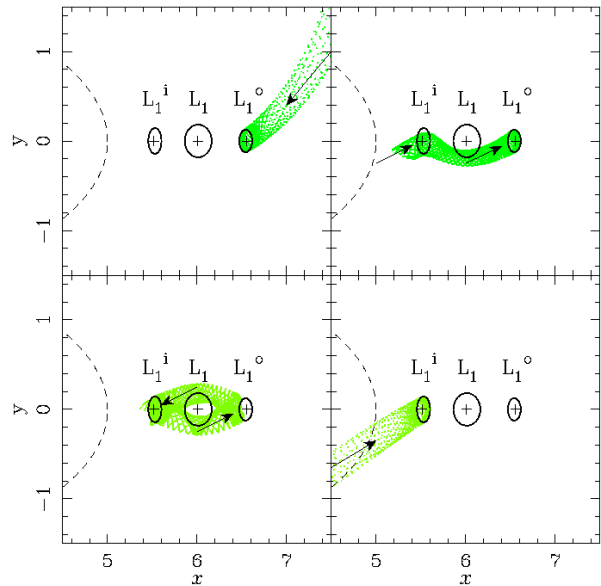}
\end{center}
\caption{Same as Fig.~\ref{fig:3Lunstable}, but for the stable
  manifolds (green). 
}
\label{fig:3Lstable}
\end{figure}

How are the dynamics in the vicinity of $L_1$ ($L_2$) modified in
cases where these Lagrangian points are stable? We will first describe
in some detail a case 
whose corresponding standard model has an $rR_1$ morphology. 
The Lagrangian points $L_{j}$, $L^i_{j}$ and $L^o_j$ ($j = 1, 2$) are
surrounded each by a family of periodic orbits. Examples of the
periodic orbits are given
with full black lines in Figs.~\ref{fig:3Lunstable} and ~\ref{fig:3Lstable}. 
The orbits in all three families are oriented perpendicular to the
direction of the bar major axis, and have shapes similar to those of the
Lyapunov orbits in the standard case. For a given energy, the orbit
around $L_{1}$ ($L_{2}$) has the largest extent, followed by that
around $L^i_1$ ($L^i_2$). The orbit around $L^o_1$ ($L^o_2$) has the
smallest extent. The orbit around $L_{1}$ ($L_{2}$) is stable while
the other two are unstable, as the Lyapunov orbits in the standard case. 

How do the invariant manifolds look in a case with six Lagrangian
points along the direction of the bar major axis?
Figs.~\ref{fig:3Lunstable} and  
\ref{fig:3Lstable} show the unstable and stable manifolds,
respectively, in the immediate neighbourhood of $L_1$. The model used
in this example is our fiducial model A (Sect.~\ref{subsec:gentheory})
with two extra Kuz'min/Toomre discs of total mass equal to
0.025$M_b$ each and a scale length $r_a$=0.6.  The unstable and the stable
manifolds  
associated with $L_1^o$ have two branches each, one ingoing and one
outgoing, as is the case for the manifolds emanating from $L_1$ in 
the standard case where $L_1$ is unstable (Paper I and
Sect.~\ref{subsec:manifolds}). The inwards unstable branch has a very
interesting morphology. Emanating from $L_1^o$, it circumvents $L_1$
from the positive $y$ values and reaches the immediate 
neighbourhood of $L_1^i$ before heading towards the outer parts of the
bar (upper right panel in Fig.~\ref{fig:3Lunstable}). The
morphology of the outgoing branch of the stable manifold of the
$L_1^o$ is also very interesting (upper right panel in
Fig.~\ref{fig:3Lstable}). Material 
moves from the bar region to the vicinity of the $L_1^i$, then
circumvents $L_1$ from negative values of $y$ to join finally the
vicinity of $L_1^o$. 

\begin{figure*}
\begin{center}
\includegraphics[scale=0.8,angle=-90.0]{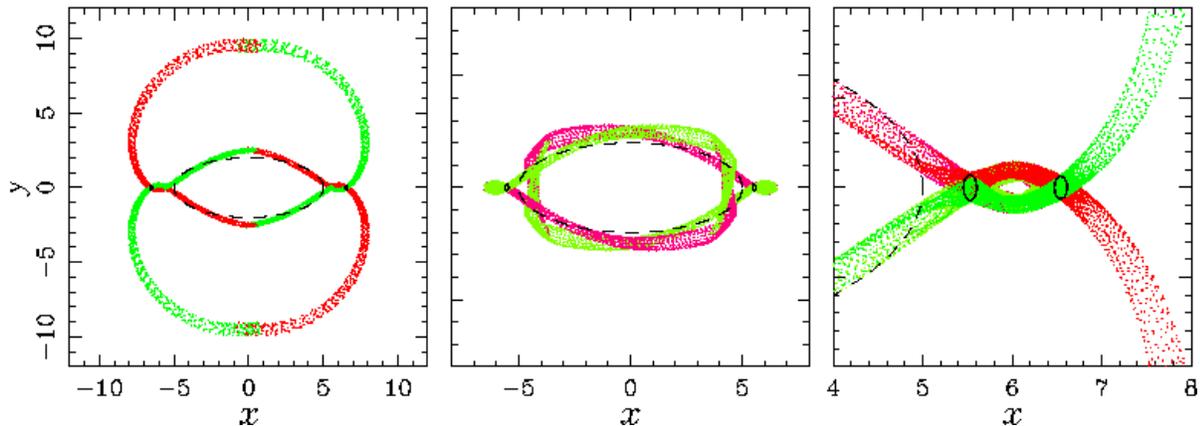}
\end{center}
\caption{Full outline of the manifolds (left panel), their inner part
  (middle panel) and the region around the Lagrangian point (right
  panel) for the model whose equipotentials are shown in
  Fig.~\ref{fig:effpot2}. Unstable manifolds are plotted in red and 
  stable ones in green. The dashed black line shows the bar outline and
  the full black lines give the periodic orbits of the same energy
  as the manifold, one around $L_1^i$ and one around $L_1^o$.
} 
\label{fig:3L1allmanif}
\end{figure*}

The manifolds of the $L_1^i$ are even more elaborate. As for $L_1^o$, the
unstable and the stable manifolds associated with $L_1^i$ have two
branches each. For the unstable manifold, one of the branches takes
material from the vicinity of $L_1^i$ and pushes it towards the outer
parts of the bar, i.e. has a very simple morphology (lower right panel
in Fig.~\ref{fig:3Lunstable}) . On the contrary, the second
unstable branch has a very complicated morphology. Material from the
vicinity of the $L_1^i$ moves outwards, circumventing
$L_1$ from the negative values of $y$ to reach the vicinity of
$L_1^o$. Before reaching it, however, it
turns around, circumvents $L_1$, now from the positive values of $y$,
and thus returns to the vicinity of $L_1^i$ (bottom left
panel of Fig.~\ref{fig:3Lunstable}). The branches of the stable
manifolds have a similar morphology (bottom panels of
Fig.~\ref{fig:3Lstable}). Note also that the innermost 
branches of the manifolds (both stable and unstable) of  
$L_1^i$, as well as the outermost branches of $L_1^o$ (both stable and
unstable) have most of their motion along the manifold, as was the case 
for the strongly unstable Lagrangian points of the standard case (see
Sect.~\ref{sec:manfprop} and Fig.~\ref{fig:weakstrman}). On the
contrary, the branches between $L_1^i$  
and $L_1^o$ are much more densely packed, with a considerable part of
the motion perpendicular to the manifold loci.

We repeated these calculations, now for a case whose corresponding
standard model (i.e. the model without the two extra Kuz'min/Toomre
discs) has a spiral structure. The dynamics in the vicinity of the
$L_1^i$, $L_1$ and $L_1^o$ 
are very similar, so we do not reproduce the corresponding plots here.
There are, nevertheless, a number of differences. One
is that now the Lyapunov orbit 
is elongated along the direction of the bar major axis,
i.e. perpendicular to the direction of the example shown in
Figs.~\ref{fig:3Lunstable} and \ref{fig:3Lstable}. Further
investigation is necessary to explain this
change of orientation. A second difference is that the morphology of
the unstable inner branch of $L_1^o$ and that of the unstable outer
branch of $L_1^i$ are reversed. Thus the former encircles $L_1$ and
returns to $L_1^o$, while the latter proceeds outwards from the
bar. The dynamics of this region needs further investigation, but we
will here be interested only in the global morphology of the manifolds
in these non-standard cases. 

The left panel of Fig.~\ref{fig:3L1allmanif} displays the full outline of
the manifolds for the example where the standard case (i.e. the case
without the two extra Kuz'min/Toomre discs) is $rR_1$ and shows a very
interesting point. Namely  
{\it the global morphology is of $rR_1$ type, the same as of
the corresponding standard case}. The loci of the branches, both stable
and unstable, of 
the $L^o_1$ and $L^i_1$ in and around the bar region nearly superpose.
A similar statement can be made for the outer branches, i.e. the ones
beyond the Lagrangian points. They, therefore, reinforce each
other. This leads to the $rR_1$ morphology,
i.e. the manifold loci form an inner ring elongated along the bar and
an outer one, 
elongated perpendicular to it, provided the distance between the     
$L_1^i$ and $L_1^o$ is not too large, i.e. in cases where the `island' of
stability around the $L_1$ ($L_2$) is of small extent. Thus the
global morphology was not changed by the stabilisation of the $L_1$
and $L_2$.

The middle panel of Fig.~\ref{fig:3L1allmanif} shows the inner parts of
the manifold clearer. We used the same model, except that we have now
integrated over longer times. Several interesting comparisons to bar
morphologies can be now 
made. First, part of the outline is rectangular-like. A second point
is that the outermost parts of this structure protrude from either
part of the bar along the direction of the bar major axis. These two
features need to be stressed here, because they will be used in Paper
IV to describe specific morphological features of observed bars.


Similarly, the model which had a spiral morphology before the two
Kuz'min/Toomre discs are added to the bar ends still keeps that
morphology after they are added and the $L_1$ and $L_2$ stabilised. 
This brings us to the interesting conclusion that the global morphology
does not change wildly, as long as the stability islands are not too
extended. Nevertheless, there are some changes, notably in the ring
diameter sizes. This will be further discussed in Paper IV, when we
compare the ratio of observed rings with the corresponding quantities
for manifolds.   

\section{The behaviour of gas}
\label{sec:schwarz}

\begin{figure}
\begin{center}
\includegraphics[scale=0.35,angle=-90.0]{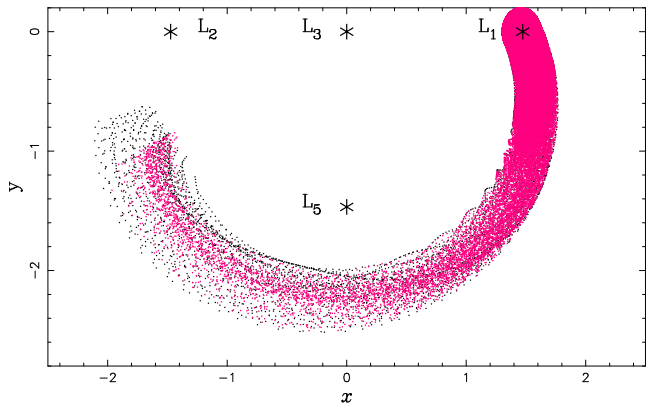}
\end{center}
\caption{Comparison of the spiral arm loci as calculated with (red) and
  without (black) collisions. See text for a description of the
  calculations. } 
\label{fig:colnocol}
\end{figure}

A number of simulations have shown the formation of spirals 
and rings from gas \citep[e.g. ][]{Schwarz81, Schwarz84,
  Schwarz85,CombesGerin85}. We should thus compare 
the dynamics of gas with that of the manifolds presented here. 
Gas, however, has different equations of motion from stars, so we need
to modify our calculations accordingly.
\citet{Schwarz79,Schwarz81,Schwarz84,Schwarz85} uses sticky particles
to simulate the gas and models 
collisions in a particularly straightforward way, so 
we can introduce a similar procedure also in our calculations. In
Schwarz's simulations, particles represent gaseous clouds which lose
a certain fraction of their kinetic energy when they collide. In
practise, they lose a certain fraction $f$ of their velocity
\citep{Schwarz81}, or only of one component of their
velocity \citep{Schwarz84,Schwarz85},
i.e. of the component of velocity that is along the line
joining the two particles. Thus, after the collision this
component is  
$v_2=-(1-f)v_1$, where the subscripts 1 and 2 designate the times
before and after the collision, respectively. The values of $f$ in
Schwarz's simulations range between 0.8 and 1. We introduced a similar
process in our calculations in the
following way. We calculated a number of orbits of the outer branch of
the unstable manifold  
and drew random numbers to find where along its trajectory each particle
will undergo a collision. We tried different numbers of collisions per
half bar 
rotation, around the values used by Schwarz. To determine the
result of a collision, we take a small box around the collision
position and calculate the average velocity of all orbits in that
box. We then decrease the relative velocity of the particle parallel to this
mean (i. e. we decrease one velocity component) by a factor $-(1 -
f)$. We ran a number of such simulations and show an example in
Fig.~\ref{fig:colnocol}. For this, we used the potential of Schwarz's standard
model for the values of the free parameters $a$ = 2.6, $q$ = 0.1 and
$\Omega_p$ = 0.27, for which he gives sufficient results and
information in his papers to allow comparisons. This potential gives
manifolds with $rR_1$ morphology. The outer branch is plotted in black
in Fig.~\ref{fig:colnocol}, together with the positions of the gaseous
particles (in red). To allow a comparison, we only include gaseous
particles that have energies above that of $L_1$. For this example
we used $f$ = 0.8, which leads to a mean energy  
dissipation per particle of 6.2 $\times 10^{-4}$, in good agreement with the
numbers given by \cite{Schwarz81}. Other numerical values give
qualitatively similar 
results. Fig.~\ref{fig:colnocol} shows that there is in general good
agreement between the loci of the gaseous and of the stellar
arm. Furthermore, the 
gaseous arm is somewhat more concentrated than the stellar arm, more
so for a larger number of collisions per revolution. In the particular
case of  Fig.~\ref{fig:colnocol} we used three collisions per half
revolution. 

Qualitatively, the above results can be understood as follows. Sticky
particles (i.e. gaseous particles) follow the same orbits as the
stellar ones, except for the collisions. This ensures a
general similarity. Due to the collisions, the gaseous particles lose
part of their kinetic energy and their velocity approaches that of the
mean. In Paper I, we compared the loci of manifolds for different
energies and found that the ones with the smaller energies lie in
configuration space {\it within} the ones with the higher
energies. This means that the corresponding spiral arms, or rings
will be thinner for the lower energies. Thus, when the particles lose
energy they will fall onto an orbit nearer to the mean and the arms,
or rings
will become thinner. This is exactly what is found with the
calculations leading to Fig.~\ref{fig:colnocol}. Thus, roughly
speaking, one can think of 
the lowest energy manifolds (i.e. the ones having the energy of $L_1$
and $L_2$) as an attractor, to which the
gaseous trajectories will tend to because of the dissipation due to
the collisions. 

\section{Summary}
\label{sec:summary}

Since the early work of B. Lindblad (\citeyear{Lindblad63}), density waves have
been commonly assumed to be at the basis of any explanation of spiral
structure formation. Here we present an alternative viewpoint, applicable
specifically to barred galaxies. This explains the formation of spiral
arms, as well as of inner and outer rings, in a common theoretical
framework. We presented this in Papers I and II and elaborate it further
here. According to our theory, it is the unstable Lagrangian points
located at the ends of the bar and the corresponding
manifolds that are responsible for the formation of 
spirals and rings. These manifolds drive orbits, which are in fact
chaotic but are confined by the manifolds, so that they create
over-densities which have the right shape to explain the spirals and
the rings. 

In Paper II we noted that different morphologies are possible, but did
not try and understand which bar properties were responsible for them.
This was one of our objectives here and to achieve it we used
three different types of bar models. In particular, we showed that 
the bar strength has a considerable effect on the properties of the
manifolds. Strong bars have more unstable $L_1$ and $L_2$ Lagrangian
points than weak bars. Material in their manifolds needs less time to
perform half a revolution around the galactic centre. The difference
can be considerable; factors of 5, or even higher, are possible. 
In the weaker bar cases, the manifolds stay
near the zero velocity curves of the same energy, while they depart
considerably from them if the bar forcing in the relevant radial
range is very strong. This brings about different morphologies. If the
non-axisymmetric forcing at and somewhat beyond $r=r_{L_1}$ is
relatively weak, the outer branches of the
manifolds have the shape of $R_1$ rings or pseudorings. Spirals, as
well as $R_2$ and $R_1R_2$ rings and pseudorings, are formed by
stronger bars.   

The circulation of material within the manifolds is also different. In
the relatively weak bar cases, which form $rR_1$ morphologies, the mass
elements move either within the inner branches, or within the outer
branches, or from the inner branches to the outer ones and then
back to inner ones again. In this third case, material is moving from the region
within corotation to the region outside it and vice versa via the
neighbourhood of the $L_1$ and $L_2$. Averaged over a sufficiently
long time, this amounts to circulation of material within a given
thick annulus, but no net motion of material inwards or
outwards. This is not true for the cases where the non-axisymmetric
forcing beyond corotation is stronger and which have a spiral
morphology. In such cases, 
material moves from the region within corotation to the region outside
it, but does not return. So, in total, this brings a net movement of
material from within corotation outwards, to the outermost parts of
the disc and may contribute to the radial extension of the disc. 

We found also that, if there is sufficient mass concentration around
the $L_1$ and $L_2$, these Lagrangian points will be stable. On either
side of each one of them and still on the direction of the bar major axis,
there will 
be another unstable Lagrangian point, so that there will be nine
Lagrangian points in total, four unstable and five stable. Two of the
stable ones are in the direction of the bar minor axis and are the $L_4$
and the $L_5$, as in the standard case. The third stable one is
located at the centre of the coordinates. The remaining
two stable ones and the four unstable ones are in the direction of
the bar major axis. This setting creates very interesting circulation
patterns around the $L_1$ and $L_2$, but leaves the global morphology
unchanged. For the case of a relatively 
weak bar, we still have an $rR_1$ morphology, but the inner and outer
rings will not necessarily touch each other. For the case of a strong
bar which had a spiral morphology before the mass concentrations around
the $L_1$ and $L_2$ were introduced, we still find spiral structure
after the mass concentrations are introduced.  

We also introduced collisions and dissipation to the manifold
calculations, in order 
to roughly model the gas properties. We found that this does not
influence the existence of the spiral arms or rings and not much their
shape and winding. The amount of dissipation, however, does influence
the width of 
the arms. These become thinner as the dissipation is increased, so that
the gaseous arm comes nearer to the lowest energy manifold. 

The next step after presenting a theory is to check whether it is
applicable to the rings and spirals observed in disc galaxies. This
will be the subject of an accompanying paper, where we will also make
a global discussion on the results of the two papers together.

\section*{Acknowledgements}

EA thanks Scott Tremaine for a stimulating discussion on the manifold
properties. We also thank Albert Bosma and Ron Buta for very useful
discussions and email exchanges on the properties of observed
rings and an anonymous referee for helpful comments. This work was
partly supported by grant ANR-06-BLAN-0172,  
by the Spanish MCyT-FEDER Grant MTM2006-00478, by a ``Becario
MAE-AECI'' to MRG, and by an ECOS/ANUIES grant M04U01.  

\appendix
\section[]{Models}
\label{app:models}

Our basic barred galaxy model in this paper will be the one introduced
by \cite{Ath92a}. Its axisymmetric component
consists of the superposition of a disc  
and a spheroid, whose basic parameters are determined so that the rotation
curve of the galactic model has the desired characteristics. The disc is modelled as 
a Kuz'min/Toomre disc \citep{Kuzmin56,Toomre63} of surface density

\begin{equation}\label{eq:kuz}
\sigma(r) = \frac{V_d^2}{2\pi r_d}\left(1+\frac{r^2}{r_d^2}\right)^{-3/2}.
\end{equation}
The parameters $V_d$ and $r_d$ set the scales of the velocities and radii, respectively. 
The spheroid is modelled using a density distribution of the form
\begin{equation}\label{eq:sph}
\rho(r)=\rho_s\left(1+\frac{r^2}{r_s^2}\right)^{-3/2},
\end{equation}
where $\rho_s$ and $r_s$ determine its central
density and scale length. Spheroids with high concentration have 
high values of $\rho_s$ and small values of $r_s$, the opposite being
true for spheroids of low concentration. Although we include two separate
axisymmetric  
components, it is important to note that, in fact, what matters in this study is only 
the total axisymmetric rotation curve and not its decomposition into components. 

The bar component is described by a Ferrers ellipsoid \citep{Ferrers77}, whose density 
distribution is described by the expression:
\begin{equation}
\rho = \left\{\begin{array}{lr}
\rho_0(1-m^2)^n & m\le 1\\
 0 & m\ge 1,
\end{array}\right.
\label{eq:Ferden}
\end{equation}

\noindent
where $m^2=x^2/a^2+y^2/b^2$. The values of $a$ and $b$
determine the shape of the bar, $a$ being the length of the semi-major
axis, which, in the rotating frame of reference, is placed along the
$x$ coordinate axis, and $b$ being the length of the semi-minor
axis. The parameter $n$ measures the degree of concentration  
of the bar. High values of $n$ correspond to a high concentration, while a 
value of $n=0$ is the extreme case of a constant density bar. The parameter 
$\rho_0$ represents the central density of the bar. For these models, the 
quadrupole moment of the bar is given by the expression
$$Q_m=M_b(a^2-b^2)/(5+2n), $$
where $M_b$ is the mass of the bar, equal to
$$ M_b=2^{(2n+3)}\pi ab^2 \rho_0 \Gamma(n+1)\Gamma(n+2)/\Gamma(2n+4) $$
and $\Gamma$ is the gamma function. 

All models with these mass components will be generically referred to in
this paper as model A. They have essentially four free parameters which 
determine the dynamics in the bar region. The axial ratio $a/b$ and the quadrupole moment 
(or mass) of the bar $Q_m$ (or $M_b$), will determine the strength of the bar. The third 
parameter is the angular velocity, or pattern speed, determined by the Lagrangian radius 
$r_L$. The last free parameter is the central density of the model 
$\rho_c=\rho_s + \rho_0$.
For reasons of continuity we will use the same 
numerical values for the model parameters as in 
\cite{Ath92a}. The axisymmetric 
component is fixed by setting a maximum disc circular velocity of
164.204 km / sec at $r$=20 kpc, and 
$r_s$ is determined by fixing the total mass of the spheroid and bar components within 
$r$ = 10 kpc to $4.87333\times 10^4~M_{\odot}$ , while fixing the combined central density of the bar and 
bulge to $\rho_c$. More information on these models can be found in
\citet{Ath92a}. 

The Ferrers bars are realistic models of bars and have been widely
used so far in orbital structure studies within and in the immediate
neighbourhood of bars 
\citep[e.g.][]{deVaucouleursFreeman72, AthBMP83, PapayannopoulosPetrou83, Pfenniger84, Ath92a, Ath92b, SkokosPA02a, SkokosPA02b}. They contain parameters with physical meaning, 
such as the bar mass or axial ratio, that can be obtained from, or compared to, observations. 
They have, however, one disadvantage, namely that in models using such bars the 
non-axisymmetric component of the force decreases very abruptly beyond a certain radius so 
that the axisymmetric component dominates in the outer regions. This is of no importance if 
one is interested in the orbital structure or the gas flow in the bar
region, as the studies mentioned above, but in studies 
like the present one, where one is interested in the region outside the bar, this may introduce a 
bias, since models with high non-axisymmetric forces beyond corotation will not
be included. 

In order to remedy this, we use here two further models, also often
used in the literature, which have  
an ad hoc bar potential, i.e. a potential that is not associated to a particular 
density distribution. Ad hoc models have some disadvantages. They are simple mathematical 
expressions for the potential and do not originate from a realistic density distribution. 
This means that the corresponding density distribution may have some undesired features, e.g. 
for very strong non-axisymmetric perturbations the total density could
even be negative locally. 
Furthermore, they do not contain simple parameters that can be
directly and straightforwardly
associated to observable quantities, like the bar length, mass, or axial ratio. Most of them 
are of the form $\epsilon A(r)\cos(2\theta)$, i.e. contain no $\cos(m\theta)$ terms with $m>2$. 
This means that the parameter $\epsilon$ is associated with the mass of the bar and that there 
is no parameter to regulate its axial ratio. Despite all these
shortcomings, ad hoc potentials 
have been widely used because they have the important advantage of being adaptable to the 
problem at hand. I.e., with a proper choice of the $A(r)$ function one can obtain a potential 
with the desired properties, for example, in our case, a potential with an important $m=2$ 
contribution between corotation and outer Lindblad resonance.

The first ad hoc bar potential we use is adapted from \cite{Dehnen00} and has the form
\begin{equation}
\Phi(r,\theta)=-\frac{1}{2}\epsilon v_0^2\cos(2\theta)\left\{{\begin{array}{ll}
\displaystyle 2-\left(r/\alpha\right)^n, & r\le \alpha\rule[-.5cm]{0cm}{1.cm}\\
\displaystyle \left(\alpha/r\right)^n, & r\ge \alpha.\rule[-.5cm]{0cm}{1.cm}
\end{array}}\right.
\label{eq:adhoc1}
\end{equation}
The parameter $\alpha$ is a characteristic length scale of the bar potential and 
$v_0$ is a circular velocity. The parameter $\epsilon$ is a free parameter related 
to the bar strength. In this paper we use $\alpha$ = 5 and
$n~=~0.75$. For this model we will use the same axisymmetric component as
for model A and we will refer to it as model D. 

Our third model has the bar potential :

\begin{equation}\label{eq:adhoc2}
\Phi(r,\theta)=\hat{\epsilon}\sqrt{r}(r_1-r)\cos(2\theta),
\end{equation}

\noindent
where $r_1$ is a characteristic scale length of the bar potential, which we will take 
for the present purposes to be equal to 20 kpc. The parameter $\hat{\epsilon}$ is related 
to the bar strength. This type of model has already been widely used in studies of bar 
dynamics \citep[e.g.][]{BarbanisWoltjer67,
  ContopoulosPapayannopoulos80, Contopoulos81}.  We will couple this
bar with the axisymmetric part used in model A and we will refer to it
as the BW model.

Throughout this paper we use the following system of units: For the mass unit 
we take a value of $10^6 M_{\odot}$, for the length unit a value of 1 kpc and 
for the velocity unit a value of 1 km/sec. Using these values, the 
unit of the Jacobi constant will be 1 km$^2$/sec$^2$.

\section[]{Linear stability analysis}
\label{app:stability}

In this appendix we will briefly present a linear solution to the
stability of the Lagrangian points. 
Setting $x_1=x,\quad x_2=y, \quad x_3=\dot{x}$ and $\quad
x_4=\dot{y}$, the equations of motion are schematically written
as a system of first order differential equations, 

\begin{equation}\label{eq-motion}
\dot{x_i}  =  f_i(x_1,\dots,x_4) ~~~~~~~~~ i = 1, \dots, 4\\
\end{equation}

\noindent
In the neighbourhood of the $L_1$ (or equivalently $L_2$) Lagrangian
points these can be written as  

\begin{equation}\label{eq-nearL1}
\left \{ \begin{array}{rclcl}
\dot{x_1} & = & f_1(x_1,\dots,x_4) & = & x_3\\
\dot{x_2} & = & f_2(x_1,\dots,x_4) & = & x_4\\
\dot{x_3} & = & f_3(x_1,\dots,x_4) & = & 2\Omega_p\, x_4-\Phi_{xx}x_1\\
\dot{x_4} & = & f_4(x_1,\dots,x_4) & = & -2\Omega_p\, x_3-\Phi_{yy}x_2.\\
\end{array} \right .
\end{equation}

\noindent
The differential matrix associated to this system is 

$$ 
Df_x(L_1)=\left ( \begin{array}{cccccc}
0 & 0 & 1 & 0 \\
0 & 0 & 0 & 1 \\
-\Phi_{xx} & 0 & 0 & 2\Omega_p \\
0 & -\Phi_{yy} & -2\Omega_p & 0 \\
\end{array} \right ).
$$
\noindent
We obtain the stability character of $L_1$ by studying the eigenvalues
of this matrix. It has four eigenvalues: $\lambda$, 
$-\lambda$, $\omega i$ and $-\omega i$, where $\lambda$ and $\omega$ are
positive values and the equations of motion can
be written as  

\begin{equation}\label{eq:sol-lingen}
\left\{
\begin{array}{l}
x(t)  =  X_1 e^{\lambda t} + X_2 e^{-\lambda t}  
        +X_3\cos(\omega t + \phi), \\
y(t)  = X_4 e^{\lambda t} + X_5 e^{-\lambda t}
        +X_6\sin(\omega t +\phi).\\
\end{array} 
\right.
\end{equation}

\noindent
If $\lambda$ is a real number, $L_1$ is a 
linearly unstable point. Larger values of $\lambda$ denote
more unstable systems. 

\bibliography{manifI_afterproofs}

\label{lastpage}

\end{document}